\documentclass[a4paper,11pt,fleqn]{article}

\usepackage[centertags]{amsmath}
\usepackage{amssymb,bm}
\usepackage{amsfonts}

\usepackage{graphicx}
\usepackage{color}
\usepackage{placeins}
\usepackage[square]{natbib}

\usepackage{float}
\usepackage{subcaption}
\usepackage{hyperref}

\usepackage{array,multirow,makecell}
\newcolumntype{R}[1]{>{\raggedleft\arraybackslash }b{#1}}
\newcolumntype{L}[1]{>{\raggedright\arraybackslash }b{#1}}
\newcolumntype{C}[1]{>{\centering\arraybackslash }b{#1}}

\newcommand{\dt}{\delta t}
\newcommand{\DT}{\Delta T}

\newcommand{\E}[1]{E\!\left[ #1 \right]}

\newcommand{\cdf}{\operatorname{cdf}}
\newcommand{\id}{\mathbb{I}}
\newcommand{\crossOverTime}{{\DT_{\!\times}}}
\newcommand{\oneYear}{ \text{1y} }
\newcommand{\var}[1]{\operatorname{var}\left[ #1 \right]}

\newcommand{\wInfty}{w_\infty}
\newcommand{\CMA}{\text{{\scriptsize CMA}}}
\newcommand{\sigmaWealth}{\sigma_\text{W}}

\newlength{\subfigureWidth}
\newcommand{\figurePannelFlat}[1]{
	\centering
	\begin{subfigure}{1.0\textwidth}
		\centering
		\includegraphics[width=\subfigureWidth]{Emp_DWE_#1}
		\includegraphics[width=\subfigureWidth]{Emp_USBond_#1}
		\caption{Historical time series: 'Developed World Equity' (left),
			'U.S. Aggregate Bonds' (right).}
	\end{subfigure}
	\begin{subfigure}{1.0\textwidth}
		\centering
		\includegraphics[width=\subfigureWidth]{M0_DWE_#1}
		\includegraphics[width=\subfigureWidth]{M0_GAB_#1}
		\caption{Simulated time series: constant drift; constant covariance; normal innovations.}
	\end{subfigure}\\
	\begin{subfigure}{1.0\textwidth}
		\centering
		\includegraphics[width=\subfigureWidth]{M1_DWE_#1}
		\includegraphics[width=\subfigureWidth]{M1_GAB_#1}
		\caption{Simulated time series: constant drift; LMARCH; NC-Student innovations.}
	\end{subfigure}\\
	\begin{subfigure}{1.0\textwidth}
		\centering
		\includegraphics[width=\subfigureWidth]{M2_DWE_#1}
		\includegraphics[width=\subfigureWidth]{M2_GAB_#1}
		\caption{Simulated time series: drift with NRC and DU; LMARCH; NC-Student innovations.}
	\end{subfigure}\\
}

\newlength{\figureWidth}
\begin{document}

\newcommand{\runtitle}{Random processes for long-term market simulations}

\begin{center}
	{ \bf \huge Random processes for long-term market simulations}

	\vspace{5ex}
	{\bf\large
		Gilles Zumbach
	}\\[2ex]
	\parbox{0.4\textwidth}{\renewcommand{\baselinestretch}{1.0}\normalsize
		Edgelab\\
		Avenue de la Rasude 5\\
		1006 Lausanne\\
		Switzerland
	}
	\\[3ex]
	\parbox{0.4\textwidth}{\renewcommand{\baselinestretch}{1.0}\normalsize
		gilles.zumbach@bluewin.ch\\gilles.zumbach@evooq.ch 

		\vspace*{4ex}
		November 10, 2025
	}


\begin{abstract}
For long term investments, model portfolios are defined at the level of indexes, a setup known as Strategic Asset Allocation (SAA).
The possible outcomes at a scale of a few decades can be obtained by Monte Carlo simulations, resulting in a probability density for the possible portfolio values at the investment horizon.
Such studies are critical for long term wealth plannings, for example in the financial component of social insurances or in accumulated capital for retirement.
The quality of the results depends on two inputs: the process used for the simulations and its parameters.
The base model is a constant drift, a constant covariance and normal innovations, as pioneered by Bachelier.
Beyond this model, this document presents in details a multivariate process that incorporate the most recent advances in the models for financial time series.
This includes  the negative correlations of the returns at a scale of a few years, the heteroskedasticity (i.e. the volatility' dynamics), and the fat tails and asymmetry for the distributions of returns.
For the parameters, the quantitative outcomes depend critically on the estimate for the drift, because this is a non random contribution acting at each time step.
Replacing the point forecast by a probabilistic forecast allows us to analyze the impact of the drift values, and then to incorporate this uncertainty in the Monte Carlo simulations.
The main change introduced by the negative return correlations is the partial decoupling between the volatility (along the time direction) from the standard deviation of the terminal values.
The definition for the process is supplemented by graphs comparing empirical results obtained from major indices with the values computed with Monte Carlo simulations.
Finally, the main statistics for the wealth at increasing time are presented, showing the key features added by the components beyond the basic normal random walk.
\end{abstract}
\end{center}
\vspace{2ex}


\newpage

\section{Introduction}
Long term financial planning is an important part of the investment activities, being for pension funds, insurances, or individual retirement plans.
Such planning involves defining a coarse-grained investment strategy using various indexes, and to check if the defined goals can be reached with the chosen strategy.
In the industry, such long term coarse-grained strategy are known as SAA, for \emph{Strategic Asset Allocation}, in opposition to the short-term detailed investment choices known as TAA, for Tactical Asset Allocation.
The time scales are quite different for both allocations, being of the order of decades for SAA, while TAA are done at the scale of days to months with actual invested positions.

SAA establishes a long-term investment plan that allocates the invested amount in different asset classes based on an investor's risk tolerance, financial goals, investment horizon, and possible other illiquid assets like real estate, private equities, or participation in businesses.
The simplest example is to decide on the balance between fixed income versus equities, hence controlling the target plan in term of risk and return.
SAA is a passive investment strategy that focuses on achieving a diversified portfolio of asset classes, that align with an investor's long-term financial objectives.
At this level, the asset classes are represented by the corresponding indexes.
For SAA, the crucial part is to check, say using Monte Carlo simulations, that the selected strategy can reach the defined goal with a chosen probability.
As an example, consider a pension fund, or an individual sparing for its retirement, with a question like: will the selected investment strategy reaches a target goal with 95\% probability.
In order to answer such questions, the distribution for the terminal wealth is needed.

The description in the previous paragraphs involves statements about the future value of an investment strategy, at some selected long time horizons.
In order to answer such questions, processes should be set, first for the evolution of the financial indexes, second for the investment strategy.
Both are quite different.
The financial world evolves according to a statistical description, at the core level as a random walk, and the investors suffer the price changes.
Very differently, the investment strategy is under control of investor(s) and manager(s), who will decide about the actual positions.
Hence, both models should incorporate these fundamental differences.
For the investment strategy, the simplest base model is a fixed-weights with periodic re-balancement, where at some predefined dates, the positions in a portfolio are bought or sold so as to be again at some defined target weights.
The periodic re-balancement is fairly simple to model and to implement, using the constraint that no in-flow or out-flow of cash is involved in the trades.
A much more difficult topic is to model the long term evolution of the financial world, and this is the subject of this paper.

The base model for market simulations follows the footsteps of \citep{Bachelier.1900} using a simple multivariate normal random walk, with constant drifts, constant covariance matrix, and normal innovations.
The parameters for this model are the drifts, volatilities and the correlation between the assets.
In the SAA context, they are collectively known as the {\em Capital Market Assumptions}, or CMA for short.
Essentially, they are long term forecasts used as input for the simulations.
The CMA are usually provided by banks, where a team of economists and data scientists reevaluate such values each year, akin to making economic forecasts at a scale of decade(s).
For the present work, the CMA are taken as given.
As shown by research on financial time series over the last 40 years, the simple model provided by Bachelier is deficient in several respects, yet is at the core of all more advanced processes.
Our goal is to incorporate in a multivariate long term model these advances, in particular the heteroskedasticity (i.e. the time dependent volatility) and the fat tails distribution for the returns.

For the empirical analyses and the Monte Carlo simulations, this paper focuses on common indexes.
Essentially, we want to simulate the long term behavior of the financial universe.
For SAA, an investment strategy is a portfolio made of positions in this universe, together with a (time dependent) allocation on these positions.
The simplest one is a buy-and-hold strategy where the initial positions are given and the portfolio is never rebalanced.
The most common strategy is using fixed-weights, where the positions are periodically rebalanced toward some target weights.
Moreover, periodic in-flows or out-flows can be used, for example modeling the contributions or consumptions of a retirement fund.
Another common strategy is a gradual shift from risky investments to fixed incomes.
Clearly, the number of interesting portfolios and strategies over a given universe is quite large, and is an interesting subject in itself but beyond the scope of this paper.
The point is that, if the dynamical and distributional properties for the indexes are correct, and if the dependencies are correctly captured, then the time series and statistics for any portfolios and strategies are also correct.
For this reason, the present empirical studies are done exclusively on indexes.

In long term studies used in SAA, the focus is often on the terminal distributions, say after 10 or 20 years.
Yet, because of the possible dynamic components in an investment strategy, it is also very important to model correctly the distributions of prices at all intermediate times.
This description must include crashes and crises, with the related high volatility.
Even though the impact of such events can average out in the long term for the indexes, this might not be the case for an investment strategy.
For example, an out-flow from a portfolio during a crash can have a large impact on its subsequent values.
For these reasons, this study attempts at best to compare empirical and simulated distributions at the longest time horizon possible, in particular in the tails.
Similarly, statistics sensitive to the dynamics of the financial time series are critical.

The core difficulty for the present study is to extract relevant information from empirical time series, in particular for long time intervals.
As an order of magnitude, at an analysis time interval $\DT$ of 1 years, 20 years of data lead to 20 independent returns.
Of course, oversampling can be used, say sampling daily or monthly the one year returns.
This will give some more information, but these returns are correlated until a lag of 1 year, and at the end, the effective sample size is still of the order of 20.
This back-of-the-envelope estimation shows the difficulty of the empirical estimations for increasing $\DT$.

The strategies that are used to extract at best information from time series are detailed in Sec.~\ref{sec:statisticalEstimators}.
Due to the lack of strong empirical statistics, we are certainly not in a comfortable position with respect to validation!
Yet, long term plannings are done anyway in the financial industry, say for pension funds or retirement plans.
Therefore, we better accept the inherent limitations in the empirical validations, and produce the best possible model given the current knowledge and historical data.
Even if imperfect, this is better than doing nothing, or using a simple normal random walk for the simulations.

To our best knowledge, many parts in the present empirical analysis and the multivariate process are new, in particular the analysis of the drift and its long term impact, the large multivariate LMARCH process, and the multivariate non-central Student.
A crucial part in long term simulations is the drifts, which are difficult to estimate.
A simple model for the drift uncertainty (DU) is introduced, taking onto account the inherent limitations of such estimations.
The uncertainty on the drift can be measured by an equivalent calibration time, essentially the length of the available historical data used to evaluate the drift.
Then, (negative) lagged correlations for the returns are added to a base constant drift, in effect stabilizing the process over long time span.
Together, these different pieces build a process suitable for long term Monte Carlo simulations of a cross-section of the financial market.

This introduction presents mainly the context and motivation for the present contribution.
Beside the historical reference to \citep{Bachelier.1900} about the process commonly used today, the author is not aware of relevant academic contributions along the general lines explained above.
The differentiation between SAA and TAA seems to be a common work path and wording in the industry, but likely without a clear academic origin.
Similarly, a large literature exist on allocations, comparing historical performances and risks, but this is not the topic of the present paper, focusing on long term multivariate processes.
Yet, several publications concern the specific components of the present process.
For these reasons, the references to previous contributions have been deported in the introduction sections for each main topics, in \ref{sec:processStructure}, \ref{subsec:drift_overview}, \ref{subsec:convariance_introduction}, \ref{subsec:randomGenerator}, where each field is summarized with the proper references.

The organization of this paper mainly follows the presentation of the introduction.
A scalar version of the process is introduced in \ref{sec:baseProcessSettings}, setting the key components and the mathematical backdrop for the process using a discrete time increment.
The scaling analysis of drift and volatility, with the impact on long term simulations, is discussed in Sec.~\ref{sec:coreScaling}.
Section \ref{sec:processStructure} introduces the multivariate structure for the processes with its main components: drift, covariance matrix, and innovations.
The technical material is presented then, namely the definitions of returns, volatilities and innovations used in this paper (Sec.~\ref{sec:quantities}), the statistical estimators used for the graphs (Sec.~\ref{sec:statisticalEstimators}), and the time series used for the empirical study (Sec.~\ref{sec:empiricalData}).
The core structure for the processes is introduced in Sec.~\ref{sec:processStructure}, the details for its components are given in section \ref{sec:drift} for the drift, in \ref{sec:covariance} for the covariance, and in \ref{sec:randomReturns} for the innovations.
These 3 sections contain graphs comparing the relevant statistics from empirical data and from Monte Carlo simulations.
After having introduced the 3 main ingredients, section \ref{sec:processesComparison} compares a selection of 5 processes in order to show their main characteristics, before the conclusions.
The more technical materials are given in appendices: appendix \ref{sec:drift_uncertainty} presents a theoretical model for the drift uncertainty, \ref{appendix:non-central-student} for the non-central multivariate Student generator, and \ref{appendix:longTermExpectations} for the long term expectation of the LMARCH process with non-central Student innovations.
The supporting statistics and graphs are given in the respective sections about each topic.

\section{The base process settings}
\label{sec:baseProcessSettings}
In order to understand the setting for the process and the scaling issue in the next section, let us first focus on the univariate case.
The setting is identical for the multivariate case, with the equations \ref{eq:process_multivariate} given below.
The base univariate process used in this paper to model the evolution of a financial index or asset is
\begin{subequations}
	\label{eq:univariate}
	\begin{align}
		r(t+\dt) & = \mu(t) + \sigma(t)\,\epsilon(t+\dt) \label{eq:r_univariate}\\
		p(t + \dt) & = p(t)\cdot \left(1 + r(t+\dt)\right). \label{eq:p_univariate}
	\end{align}
\end{subequations}
In these equations, the current time is $t$, with the past information known up to and including $t$.
This means that the drift $\mu(t)$ and the volatility $\sigma(t)$ can be evaluated using the information up to $t$ (i.e. they are in the information set, or are in the filtration $\mathcal{F}(t)$, or are $t$-measurable).
The innovations $\epsilon(t+\dt)$ is unknown at $t$, independent from the previous draws, and described by a random variable with a distribution $p_\epsilon(0,1)$.
This distribution has a zero mean and unit variance.
Importantly, $p_\epsilon$ is stationary, namely the distribution's shape is fixed.
The equation \ref{eq:r_univariate} specifies the (random) return over one time step $\dt$, Eq.~\ref{eq:p_univariate} the time evolution of the (random) prices given the return.
As seen from $t$, $\epsilon(t+\dt)$, $r(t+\dt)$ and $p(t + \dt)$ are random variables.
At $t+\dt$, the realized price $p(t + \dt)$ becomes known, from which $r(t+\dt)$ and $\epsilon(t+\dt)$ can be evaluated by inverting the above equations, namely all become part of the filtration $\mathcal{F}(t+\dt)$.

The point of view used in this paper is of a random process with a discrete time increment $\dt$.
Because of the distribution $p_\epsilon$ has fat tailed in order to be realistic, and because of the drift and volatility dynamics provided by $\mu(t)$ and $\sigma(t)$, the present specifications are unlikely to have a continuum limit and to be described by some Ito stochastic differential equations.
Hence, only a discrete time formulation of the process is used.

Both equations \ref{eq:univariate} or \ref{eq:process_multivariate} can be used directly in Monte Carlo simulations for applications in finance.
As stated in the introduction, our goal is to have realistic long term processes, matching the empirical properties of the financial markets.
Rigorous mathematical questions are not studied, say like the existence of asymptotic long term distributions for the return or the variance.
For the simulations used to produce the figures, a time step $\dt$ of 1 month was used.
Yet, the equations are not rooted at a particular value for the time increment, and a step of 1 day or 1 year could be used, depending on the application.

\section{The core scaling of drift and volatility}
\label{sec:coreScaling}
For the sake of the present scaling argument, let us assume that the drift and volatility have no time dependency, namely are real numbers, say equal to their long term values.
These parameters are in general given at an annual scale, for example the typical annualized volatility of a stock index is between 15 to 30\%.
Yet, the process is formulated at the scale $\dt$, and the parameters have to be scaled to $\dt$ such that the resulting random walk has the correct behavior at longer time scale, say for example the specified annualized volatility.
Intuitively, the drift and volatility parameters must decrease with $\dt$ such that the drift and volatility at 1 year correspond to the desired values.
In the equation \ref{eq:r_univariate}, $\mu$ and $\sigma$ are respectively of order $\dt$ and $\sqrt{\dt}$, originating in the deterministic and in the diffusive components of a random walk.
The $\dt$ dependency can be made explicit by using annualized parameters, with
\begin{subequations}
	\label{def:scaling}
	\begin{align}
		\mu = & \mu_{\dt}  = \mu_\oneYear \,\frac{\dt}{\oneYear} \label{def:scaling_return}\\
		\sigma = & \sigma_{\dt} = \sigma_\oneYear \,\sqrt{\frac{\dt}{\oneYear}} \label{def:scaling_volatility}
	\end{align}
\end{subequations}
where ``1y'' is a one year time interval, the subscript ``1y'' denotes annualized parameters, and the subscript ``$\dt$'' makes explicit the implicit time scale in the drift and volatility.
The ratio $\dt/\oneYear$ is the process time step expressed in year.
The same relation can be used for any time interval $\DT$, showing that the drift grows as $\DT$ while the diffusive part grows as $\sqrt{\DT}$, as usual for a random walk.

These different scalings for the drift and volatility are crucial to understand the dominant behavior of a random walk, and the long time problems in finance.
For small $\DT$, say from days to a few years, the volatility dominates the drift, and the random component is the leading feature of the process.
This is the regime where most computations are done in finance, for example in portfolio optimization, risk evaluation, or option pricing.
For large $\DT$, say above one decade, the drift dominates the random part, and this deterministic component is the leading feature of the process.
The cross-over from diffusion to drift occurs at a time $\crossOverTime$ when
\begin{equation}
	\mu_\oneYear \frac{\crossOverTime}{\oneYear} - \sigma_\oneYear \sqrt{\frac{\crossOverTime}{\oneYear}} = 0.
	\label{eq:eq_for_cross_over_time}
\end{equation}
In term of the probability distribution of the return, this equation says that the center of the distribution (the term in $\mu_\oneYear$) is one sigma away  (the term in $\sigma_\oneYear$) from 0.
Essentially, after the time $\crossOverTime$, the drift is large enough to become visible with a good probability.
The equation \ref{eq:eq_for_cross_over_time} can be solved for $\crossOverTime$
\begin{equation}
	\crossOverTime = \left(\frac{\sigma_\oneYear}{\mu_\oneYear}\right)^2 \,\oneYear
\end{equation}
where $\crossOverTime$ is expressed in year.
Notice that $\mu_\oneYear$ appears in the denominator.
The more familiar Sharpe ratio is closely related with the cross-over time
\begin{equation}
	r_\text{Sharpe} = \frac{\mu_\oneYear}{\sigma_\oneYear} = \sqrt{\frac{\oneYear}{\crossOverTime}}.
\end{equation}
When estimated from empirical time series, the drift is depending only on the start and end values of the sample, therefore carry a large uncertainty.
The mean volatility appears to have better estimation properties, at least for constant volatility process.
Yet, the heteroskedasticity makes it also dependent from the sample, roughly with a very large crisis each 10 years.
Therefore, the empirical estimation of both $\mu$ and $\sigma$ have important dependencies on the available empirical sample.
Hence, the cross-over time and the Sharpe ratio are ``fragile'' with respect to the empirical estimation of $\mu$ and $\sigma$, and should be considered only as an order of magnitude.

Using long spans of historical data, the drift, volatility and cross-over time can be evaluated for a few indexes, as reported in the  table~\ref{table:crossoverTime}.
Roughly, the cross-over time is of the order of months to 1 decade for fixed income indexes, and around years to a few decades for equity indexes.
These numbers are in line with the typical investment advice, namely fixed income investments should be used for low risk and short investment periods, while equities are more risky and should be considered for long investment periods.
A balanced portfolio allows to adjust the investment strategy to the desired risk and return profile.

\begin{table}[ht]
	{\small
		\begin{tabular}{l|lrrrr}
			name  &  category  &  return  &   volatility  &  $\crossOverTime$  &  Sharpe ratio  \\
			\hline
			U.S. Short Duration Government &  FI  & 0.021 & 0.014 & 0.42   & 1.53  \\
			Chinese Government Bonds    &  FI     & 0.048 & 0.038 & 0.64   & 1.25  \\
			U.S. Aggregate Bonds        &  FI     & 0.032 & 0.042 & 1.77   & 0.75  \\
			U.S. High Yield Bonds       &  FI     & 0.065 & 0.094 & 2.10   & 0.69  \\
			U.S. Inv Grade Corporate Bonds & FI   & 0.042 & 0.066 & 2.45   & 0.64  \\
			U.S. Government Bond        &  FI     & 0.025 & 0.045 & 3.15   & 0.56  \\
			Emerging Markets Sovereign Debt & FI  & 0.049 & 0.094 & 3.65   & 0.52  \\
			Euro High Yield Bonds       &  FI     & 0.057 & 0.157 & 7.71   & 0.36  \\
			U.S. Long (20+ Yr) Treasuries & FI    & 0.037 & 0.140 & 13.96  & 0.27  \\
			World Government Bonds      &  FI     & 0.017 & 0.067 & 16.06  & 0.25  \\
			Euro Inv Grade Corp Bonds   &  FI     & 0.021 & 0.111 & 27.35  & 0.19 \\
			Canadian Large Cap          &  Equity  & 0.082 & 0.119 & 2.10  & 0.69  \\
			U.S. Large Cap              &  Equity  & 0.098 & 0.155 & 2.51  & 0.63  \\
			U.S. Small Cap              &  Equity  & 0.073 & 0.204 & 7.83  & 0.36  \\
			European Small Cap          &  Equity  & 0.059 & 0.223 & 14.20 & 0.26  \\
			European Large Cap          &  Equity  & 0.043 & 0.188 & 18.75 & 0.23  \\
			MSCI China Equity           &  Equity  & 0.053 & 0.259 & 24.09 & 0.20  \\
			UK Small Cap                &  Equity  & 0.045 & 0.225 & 24.74 & 0.20  \\
			UK Large Cap                &  Equity  & 0.035 & 0.175 & 24.83 & 0.20  \\
			Emerging Markets Equity     &  Equity  & 0.040 & 0.210 & 28.03 & 0.19  \\
			Japanese Equity             &  Equity  & 0.027 & 0.151 & 31.23 & 0.18  \\
			Relative Value Hedge Funds  &  Alt.    & 0.051 & 0.049 & 0.93  & 1.04 \\
			Event Driven Hedge Funds    &  Alt.    & 0.051 & 0.070 & 1.90  & 0.72 \\
			Macro Hedge Funds           &  Alt.    & 0.033 & 0.048 & 2.07  & 0.70  \\
			Conservative Hedge Funds    &  Alt.    & 0.027 & 0.041 & 2.29  & 0.66  \\
			Diversified Hedge Funds     &  Alt.    & 0.028 & 0.050 & 3.11  & 0.57  \\
			Long Bias Hedge Funds       &  Alt.    & 0.046 & 0.089 & 3.67  & 0.52  \\
			Global Core Infrastructure  &  Alt.    & 0.021 & 0.162 & 59.30 & 0.13
		\end{tabular}
	}
	\caption{Mean return, mean volatility, cross-over time $\crossOverTime$ (in year) and Sharpe ratio for some financial assets.
		All values are computed in USD, over the period Jan 2006 to Dec 2023, with monthly returns, then annualized.
		The values are ordered according to the asset type (Fixed Income, Equity and Alternative), then by increasing cross-over time.
	}
	\label{table:crossoverTime}
\end{table}

In all these scaling arguments, a risk free rate $r_\text{risk-free}$ can be subtracted from the drift or be inserted explicitly in the equations, depending on the reader preferences and the domain of application.
In view of the small values for the short term interest rates over the last two decades (between negative to 1\%), including $r_\text{risk-free}$ does not change the core of the present order of magnitudes.

Returning to the problem of long term market simulations, the key point is that \emph{the drift is the critical parameter} since the simulation time is at a scale of decades, namely above the cross-over time.
This setting is fundamentally different from most computations done in finance where the drift is a second order correction to results dominated by $\sigma$ (say for example in the Black-Sholes option pricing formula, or in market risk evaluation).
For this reason, the drift is an important part of this paper, and includes a dynamical component to introduce a mean reversion, and a study of the error on the final distribution induced by the numerical uncertainty on the mean drift parameters (see Sec.~\ref{sec:drift}).

\FloatBarrier
\section{The multivariate process structure}
\label{sec:processStructure}
The multivariate version of Eq.~\ref{eq:univariate} is
\begin{subequations}
	\label{eq:process_multivariate}
	\begin{align}
		\bm{r}(t+\dt) & = \bm{\mu}(t) + \left(\bm{\Sigma}(t)\right)^{1/2}\,\bm{\epsilon}(t+\dt) \label{eq:r_multivariate}\\
		\bm{p}(t + \dt) & = \bm{p}(t)\odot \left(1 + \bm{r}(t+\dt)\right)  \label{eq:p_value}
	\end{align}
\end{subequations}
where vectors and matrices are denoted with bold-face characters.
The prices $\bm{p}$ follow a random walk driven by the relative returns, with the same process setting as for Eq.~\ref{eq:univariate}, namely with a finite time increment $\dt$.
Beware that in Eq.~\ref{eq:p_value}, the product denoted by $\odot$ in the rhs is an element-wise product (known as the Hadamard product).
The volatility matrix used in the process is a square root of the covariance matrix $\bm{\Sigma}$.
Eq.~\ref{eq:r_multivariate} is the specification for the return process used in this paper, and where the drift vector $\bm{\mu}(t)$, covariance matrix $\bm{\Sigma}(t)$, and innovation distribution $p_{\bm{\epsilon}}$ should be specified.
As for the univariate case, the multivariate distribution $p_{\bm{\epsilon}}$ is assumed to be stationary, with a zero mean and unit variance, but is not assumed to be normal.
With this structure, all the dependencies from the past are summarized in $\bm{\mu}(t)$ and $\bm{\Sigma}(t)$.
In particular, the vast family of ARCH model falls into this framework.
An important task is to justify empirically these dynamical components.

The simplest model is the Bachelier process, with constant drift and volatility, and a normal distribution for the innovations, with the multivariate parameters:
\begin{itemize}
	\item the vector of drift $\bm{\mu}$,
	\item the vector of volatility $\bm{\sigma}$, and
	\item the matrix of correlation $\bm{\rho}$.
\end{itemize}
With these parameters, the covariance is given by
\begin{equation}
	\bm{\Sigma} = \bm{\sigma_D} \cdot \bm{\rho} \cdot \bm{\sigma_D}
\end{equation}
and $\bm{\sigma_D}$ is the diagonal matrix containing the volatility for the assets given in $\bm{\sigma}$.
This set of parameters are known as CMA, and are forecasts for the corresponding values.
Some long term econometric models can be used to derive them, mainly the drift $\bm{\mu}$ and volatility $\bm{\sigma}$.
As a simple alternative, all CMA parameters can be evaluated with historical data (using an implicit stationary assumption).
For the present work, the CMA are taken as granted.

The CMA parameters have some constraints.
There is no constraint on the drifts, and the volatilities must be positive.
The most important constraint is on the correlation matrix, which must be a definite positive matrix.
This limitation originates in the process equation, where the square root of the covariance matrix must be taken\footnote{
	Mathematically, the correlation matrix must be non negative, namely all eigenvalues obey $\epsilon_i \geq 0$.
	Because the square root will be computed numerically, the eigenvalues must be above zero within some tolerance $\epsilon_i \geq \epsilon_\text{min} > 0$ in order to ensure that, numerically, the square root can always be evaluated.
}.
This constraint on a matrix is fairly limiting, and prevent to mangle with the correlation elements (say for example that some selected correlation values are projected to be larger).
Essentially, the correlation matrix must be computed from past returns, and with the same formula and time span for all time series.
An alternative approach is to project in some sense a non-positive correlation matrix on a space of positive correlation matrices, see for example \citep{Ortec.CovarianceCompletion} for a recent work in this direction.

On top of the constant CMA parameters, dynamics can be added on $\bm{\mu}(t)$ and $\bm{\Sigma}(t)$, and the probability distribution for the innovations must be specified.
Since these parts are independent, a brief summary is given here, while a specific introduction, the analytical formula and validation statistics are presented respectively in sections \ref{sec:drift}, \ref{sec:covariance} and \ref{sec:randomReturns} respectively.

As mentioned above, the drift is the critical parameter in this long time problem.
Large error bars should be assumed on the CMA drift, and an error of order $\delta \mu$ on the drift has an impact $\DT \cdot\delta \mu$ after a time $\DT$.
This behavior is dominant compare to the diffusive term, which grows as the square root of $\DT$, hence it is important to include such uncertainties in the long term projections.
The simplest \emph{drift uncertainty} (DU) model is to assume random drifts with a normal distribution centered around the CMA value.

Beside, lagged correlations for the returns and innovations point to positive correlations up to a few months, and negative correlations around a few years.
Such statistics are in agreement with short term ``trend following'' of traders, and with a ``mean reversion'' at longer time scales.
This behavior is embedded in the drift dynamics $\bm{\mu}(t)$ by lagged return terms, generating similar statistics in Monte Carlo simulations.
The most important term is the negative lagged correlations, which introduce a correcting effect after large drops or rises in the simulated prices, effectively ``stabilizing'' the long term process.
We have called this term \emph{Negative Return Correlation} (NRC).

The model used for the drift is detailed in Sec.~\ref{sec:drift}, with a specific overview \ref{subsec:drift_overview}, the Monte Carlo simulations is explained in \ref{subsec:DU} (DU) and \ref{subsec:NRC} (NRC), the empirical statistics on the drifts are presented in Sec.~\ref{subsec:empirical_validation_NRC}, and the analytical model for the drift uncertainty is given in the Appendix~\ref{sec:drift_uncertainty}.

The volatility dynamics is another important component of a process used to describe the time evolution of financial assets.
Since the seminal works of Engle and Bollerslev on ARCH processes \citep{Engle.1982, Bollerslev.1986}, the heteroskedasticity, namely the non-constant volatility of the empirical data, has been recognized as an important stylized fact.
The extensive family of ARCH processes can model the changing volatility that occur in the financial time series.
Among them, the Long-Memory ARCH process (LMARCH) is a very good workhorse for many applications in finance \citep{Zumbach.LongMemory, Zumbach.book}.
This process is fairly sparse, has a few parameters which do not need to be estimated on empirical data, and allows to compute volatility forecasts for any desired horizons.
In short, this process is a robust base  model to describe the heteroskedasticity present in financial time series.

Our goal is to describe a multivariate universe of correlated indexes.
This raises significantly the challenge since a multivariate ARCH process is needed, a difficult task.
A key issue is the increasing number of parameters with the universe size $n$, since the proposed extensions have a number of parameters growing as $n^2$, or even $n^4$, making them untractable even for moderate universe size.
Hence, the challenge is to write a simple but good enough multivariate ARCH process.
For this purpose, we leverage the good model of heteroskedasticity for all financial time series provided by the LMARCH process \emph{with the same parameters}.
Essentially, the multivariate process is a cross-product of one univariate LMARCH process.
The number of parameters is given by the asymptotic long term covariance matrix, namely $n$ asset volatilities and $n(n-1)/2$ correlations (all provided in the CMA), and a fixed number of parameters to describe the memory decay of the LMARCH process.
The multivariate dynamical model used for the covariance is described in Sec.~\ref{sec:covariance},
with an overview on this topic (\ref{subsec:convariance_introduction}), the constant covariance model provided by the CMA (\ref{subsec:convariance_constantCovariance}), the LMARCH extension (\ref{subsec:lmarchProcess}),
and the statistics for the volatility in \ref{subsec:vol-on-vol_empirical}.
The asymptotic property of the mean covariance is verified in Appendix~\ref{appendix:longTermExpectations}.

The last component in the process is the multivariate random generator for the innovations.
The default solution is to use a normal generator, with covariance $\Sigma$.
Yet, as shown by many empirical studies and by the investigations in Sec.~\ref{sec:randomReturns} below, the distributions are clearly fat-tailed.
Moreover, a systematic asymmetry is observed, with larger down moves than up moves.
This asymmetry is observed for stock indexes (larger crashes than rallies), but also for fixed income indexes (more abrupt interest rate increases than decreases).
These observations call for an asymmetric fat-tailed distribution, and the univariate non-central Student distribution provides for a simple and good description of the empirical data.
For Monte Carlo simulations, a multivariate generator is needed, with a specified mean $\bm{\mu}$ and covariance $\bm{\Sigma}$.
\citep{Embrechts.2015} gives an algorithm to generate multivariate non-central Student variates, but the mean and covariance cannot be specified simply.
This algorithm has been reformulated so that the mean and covariance are as specified.
The random generator issues are presented in (\ref{subsec:randomGenerator}), the analyses of the empirical and simulated distributions for the returns and innovations are provided in Sec.~\ref{subsec:innovation_distributions} and \ref{subsec:return_distributions}, the random generator algorithm is detailed in Appendix~\ref{appendix:non-central-student}.

Finally, 2 points need to be discussed with respect to Eq.~\ref{eq:process_multivariate}.
First, as such, the prices do not necessarily stay positive.
An absorbing state at zero is added, with $p(t+\dt)$ replaced by zero when $p(t+\dt) \leq p_\text{min}$.
The parameter $p_\text{min}$ can be set at a small enough value, appropriate for each index, say 1.
Since the intention is to model broad indexes, reaching the absorbing state means the bankruptcy of a state or an economy, an unlikely event.
At the start of a simulation, the current price $p(0)$ is known, and the minimal price is set at $p(0)/100$.
If reaching such events are too likely in a simulation, this denotes that the model or the model's parameters are inappropriate.
This consideration has lead to the introduction of the NRC term (negative return correlation) in order to stabilize the long term diffusion.
Yet, if this model is used for stocks, bankruptcy is a possible outcome and must be properly included.

Second, the simulation universe can be over several currencies, and FX rates need to be included.
This can be done in 2 ways.
If a reference currency is given, all indexes can be converted to the reference currencies, and the CMA parameters are including the exchange rates impacts.
This is the simplest solution with respect to the applications, since the FX rate effects are included in the CMA, but the CMA become dependent on the reference currency.
The second solution is to include explicitly the FX rates in the simulation universe, and to do the FX conversions where appropriate.
In this case, the CMA for the indexes are given in their respective home currency, but the simulation universe and the CMA have to be extended for the FX rates.
As an example, consider a fixed income index in EUR with a 1\% (annualized) volatility, included in a USD portfolio.
The EUR/USD fx rate has a typical volatility of 8\%.
With the first solution, the volatility of the fixed income EUR index is computed in USD, with a resulting volatility around 8\% due to the FX fluctuations, while with the second solution, the FX rate is included explicitly and each components retain its volatility.
The first solution is simpler at the level of the implementation for the process and subsequent portfolio evaluation, the second is simpler for the evaluation of the CMA.
Yet, the equations \ref{eq:process_multivariate} remains unchanged for both solutions.
Beside, with respect to FX, all the empirical statistics reported in this paper have been evaluated in their home currency.

\section{Definitions of historical return, volatility and innovation}
\label{sec:quantities}
The time interval of interest is denoted by $\DT$, the time step for the historical data or the simulation process is $\dt$.
The current time is $t$, and the information is available up to this time (i.e. the filtration is $\mathcal{F}(t)$).
With a time series of prices $p(t)$, either historical or simulated, the (realized) relative return is defined as
\begin{equation}
	\label{def:realizedReturn}
	r(t; \DT) = \frac{p(t) - p(t-\DT)}{p(t-\DT)}
\end{equation}
This definition is equivalent to Eq.~\ref{eq:p_univariate}, but at the scale $\DT$.
The resulting returns are at the scale $\DT$, they can be annualized using the scaling $r_\oneYear = r_{\DT} \sqrt{\oneYear/\DT}$.

The forecast for the variance $\sigma^2$ over the interval $[t, t+\DT]$ is computed as
\begin{equation}
	\label{def:varianceForecast}
	\sigma^2(t; \DT) = \sum_{t' \leq t} w( t-t', \DT) \,r^2(t'; \dt) = \sum_{0 \leq l \leq l_\text{max}} w( l, \DT) \,r^2(t - l\,\dt; \dt)
\end{equation}
where the sum runs over $t'$ in the past of $t$, namely $t' = t - l\,\dt$, or equivalently over the lag $l$.
The volatility forecast is at the scale $\dt$, a pre-factor can be included in order to scale the variance at the scale $\DT$ or at 1 year.
The weights $w$ sum to 1 regardless of $\DT$
\begin{equation}
	\label{eq:sumW}
	\sum_{t' \leq t} w(t-t', \DT) = \sum_{l \geq 0} w(l, \DT) = 1.
\end{equation}
Essentially, the variance forecast at the horizon $\DT$ is a sum of the past squared returns at the scale $\dt$, with weights depending on the distance to the present time $t$, and on the forecast horizon $\DT$.
The decay of the past information is set by $w(l, \DT)$ with $l$ the lag from the present time $t$.
Different weights $w$ lead to various volatility estimators, and 2 estimators are used for the empirical analyses.

The first estimator is the RMA, for Rectangular Moving Average, with constant weights in a window of length $\DT$, namely $w(l, \DT) = \dt/\DT$ for $0 \leq l < \DT$ and zero otherwise.
With this definition, $\sigma(t; \DT)$ computes the realized volatility over the interval $(t - \DT, t)$, and does not use information in the past of $t-\DT$.

The second estimator is derived from the Long-Memory ARCH process.
This volatility process specifies the 1 step volatility weights $w(l, \dt)$ with a slow decay with respect to the lag $l$, capturing the slow decay of the information as the return $r(t - l\,\dt; \dt)$ recedes in the distant past with increasing lag $l$.
Because of its simple quadratic structure, expectations of the variance between $t$ and $t+\DT$ can be evaluated analytically conditional on the information available up to $t$. 
This expectation allows to compute variance forecasts for any horizon $\DT$ with the formula \ref{def:varianceForecast} (rigorously, the forecasts should carry a tilde, and a volatility forecast is the square root of a variance forecast).
The weights $w(l, \DT)$ are computed from the 1-step weights $w(l, \dt)$ with a recursive equation.
The key point is that no new parameters are needed, namely the volatility forecast for the horizon $\DT$ is obtain from the 1-step forecast.
Moreover, the same parametric weights $w(l, \dt)$ describe correctly most free-floating financial assets, regardless of the asset type (indexes, equities, fixed incomes) and geographic area \citep{Zumbach.RM2006_fullReport}.
In short, the LMARCH model is a good default model, that can be used for any time series, and without estimating specific parameters.
It is not the best model in store, but it is very efficient in term of capturing the heteroskedasticity, namely crashes versus quiet periods, and with a low complexity.

Eq.~\ref{eq:r_univariate} is interesting to analyze historical time series because it allows us to define the realized innovations for one asset with
\begin{equation}
	\label{def:innovations_dt}
	\epsilon(t+\dt) = \frac{r(t+\dt) - \mu(t; \dt)}{\sigma(t; \dt)}.
\end{equation}
Essentially, this formula ``de-GARCH'' the returns.
In the process definition, it is assumed that the distribution $p_\epsilon$ is stationary.
The equivalent statement on historical data is that Eq.~\ref{def:innovations_dt} should lead to a stationary time series, at least to a good approximation.
The formula \ref{def:innovations_dt} is obtained from a process with a time step $\dt$.
The same formula can be used at a scale $\DT$ to define $\epsilon(t+\DT)$,
where $\sigma$ is a volatility forecast for the horizon $\DT$ and computed at $t$.
Essentially, this formula removes from $r(t+\DT)$ the information known at $t$ in term of the location $\mu(t; \DT)$ and size $\sigma(t; \DT)$, in order to measure the ``surprise'' $\epsilon(t+\DT)$ over the interval $\DT$.

From the empirical statistical analysis, long term asymptotic values are desired.
Usually in finance, the historical returns are used, say for example to estimate the mean variance.
Yet, the returns have a changing variance, namely the return are heteroskedastic, with very large crises recurring approximately each decade (for stocks over the last 30 years: dot.com bubble, subprime crisis, covid).
Hence, an asymptotic estimate for the variance requires a couple of decades of data.
The situation is similar with other estimates, say the return distribution and its tails, or lagged correlations.
Because stationary, the statistical properties of the innovations converge faster to their asymptotic (long sample) limits.
This is particularly interesting for studies at long time horizons, because the effective sample size is always small for large $\DT$.

\section{Statistical estimators}
\label{sec:statisticalEstimators}

\subsection{General context}
Because of the poor samples for long term statistics, it is important to select good statistical estimators, for the distributions and for the dynamical properties.
Several strategies are used to extract at best information from historical data.
\begin{itemize}
	\item Obviously, use long time series, but this is limited by the availability of data.
	Another limit is the long term stability of the economy and the financial system, say at the scale of several decades.

	\item Use a cross-section of indexes, for equities and fixed incomes, and with a good geographical coverage.
	The purpose here is to obtain \emph{universal properties}, or ``stylized facts'', namely statistics occurring similarly across indexes, eventually depending on the asset class (e.g.equity indexes versus fixed income indexes).
	Such properties allow to build universal and robust model for the process, by focusing only on the dominant behaviors.
	Yet, the correlations between indexes limit the diversification provided by large cross-sections, in particular for equity indexes.

	\item Use efficient statistical estimators.
	Two statistics are used in this work.
	First, distributional information is needed, and \emph{folded-cdf} are used to display the core and both tails of various quantities.
	Second, ``\emph{lag-one correlation}'' statistics at increasing $\DT$ are used to obtain information about the dynamics.
	Both statistics are explained below.

	\item Extract information about several quantities, in particular returns, volatilities and innovations.
	Sec.~\ref{sec:quantities} gives the definitions used in this work.

	\item Extract information on increasing time intervals, ranging from 1 month to a few years.
	In line with the effective sample size argument presented in the introduction, the statistical uncertainties grow with the analysis time interval $\DT$.
	More confidence is build when a consistent picture emerges, when increasing $\DT$ and cross-sectionally.

	\item Take care of the finite sample bias.
	Because the computed statistics are never in a large sample regime, important finite sample bias are present.
	Monte Carlo simulations with a base normal random walk allow to estimate the bias.

\end{itemize}

\subsection{Folded-cdf}
\label{subsec:foldedCdf}
Given a sample of data, either from historical data or Monte Carlo simulations, the cdf (cumulative distribution function) can be evaluated simply by a counting procedure.
This contrast with a pdf where a density must be evaluated with some kernel, which is distorting the empirical data.
For the graphical representation, we plot the ``folded-cdf'', or f-cdf, namely $\cdf(x)$ for $\cdf \leq$ 1/2, and $1 - \cdf(x)$ for $\cdf$ > 1/2.
With this representation of f-cdf($x$) versus $x$, a vertical logarithmic axis allows to show both tails of the cdf, together with the core of the distribution.
Cumulative probabilities are shown directly on the graph, allowing for a direct visualization of VaR values.
The cdf's are computed for increasing $\DT$ on annualized data, in order to remove the trivial random walk scaling.
Showing on the same graph the folded-cdf for increasing $\DT$ emphasizes the similarities and changes in the distributions.
On some f-cdf graph, the cdf of a normal distribution as been added for comparison, represented with a dashed black line.
Its mean and standard deviation are the empirical moment estimators for the 1 month data.

\subsection{Lag-one correlations}
\label{subsec:lagOneCorrelation}
For two quantities $X$ and $Y$ measured at a time horizon $\DT$, the correlation between $X(t)$ and $Y(t + \DT)$ gives the largest information contained in $X$ about the subsequent evolution of $Y$ at the scale $\DT$.
The time displacement in $Y$ is chosen such that there is no trivial overlap in the underlying prices or returns for the computation of $X$ and $Y$.
For two quantities $X$ and $Y$, we call the ``lag-one $\DT$ correlations'' the lagged correlations at one-$\DT$ as function of $\DT$.
The return-return lag-one correlations (Fig.~\ref{fig:lagCorr_scaling_equities} to \ref{fig:LaggedCorrelation_returnHist_Vs_returnReal}) are particularly important for the present analysis, since this modifies the long term dispersion of the prices.
Notice that this is different from the usual approach of computing $X$ and $Y$ at the smallest scale $\dt$, then to compute the lagged correlations for increasing lag $k\,\dt$.
The variable is the increasing lag in the usual approach, whereas the increasing value $\DT$ is used as the parameter in our approach, a parameter that impacts both the quantities $X$ and $Y$, and the lag.
Because this maximizes the information about $X$ on the subsequent $Y$ for all scales $\DT$, there is no need to evaluate correlations for larger lags, say $k\,\DT$ for $k > 1$.

The lag-one correlation for the volatility is also very important since measuring the heteroskedasticity, namely the clustering of the volatility as function of $\DT$.
When using $\sigma_\text{LMARCH}[\DT]$ at $t$ and $t+\DT$, the long memory kernel creates a spurious dependency since some squared returns are common to both volatilities.
Furthermore, the volatility has a clearly skewed distribution, whereas the logarithmic volatility shows a more symmetric distribution.
In order to avoid both issues, the heteroskedasticity is best quantified with $X = \log(\sigma_\text{LMARCH}[\DT])$ (with $\sigma_\text{LMARCH}[\DT]$ a volatility forecast for a time horizon $\DT$ computed with an LMARCH process), and $Y = \log(\sigma_\text{RMA}[\DT])$ (a volatility computed with equal weights over a span $\DT$ of monthly squared returns).

For Monte Carlo simulations, the lag-one correlations $\rho_{X, Y}(\DT)$ are computed for each path.
Then, the average $m_\text{MC, X, Y}$ and standard deviation $\sigma_\text{MC, X, Y}$ of  $\rho_{X, Y}$ are evaluated over the paths.
On the graph for the lag-one correlation obtained from Monte Carlo simulations, see figures \ref{fig:LaggedCorrelation_returnHist_Vs_returnReal} and \ref{fig:LaggedCorrelation_logVol_lmarchHist_Vs_logVol_rmaReal}, the average $m_\text{MC}$ is plotted with a black line.
Two red lines are added at a distance of $\pm 1.95 \sigma_\text{MC}$ to denote the 95\% confidence bounds for this quantity.
Since the Monte Carlo simulations are done for the same time length as the empirical data, a similar confidence bound can be assumed for the empirical graphs.

\section{Empirical data}
\label{sec:empiricalData}

A good model should reproduce at best the statistical properties of financial markets.
A validation must be done comparing some statistics computed with market prices and with Monte Carlo simulations.
The difficulties are on the market data side, since an order of 20 years of historical data are available with a large market cross-section.
At the scale of 1 month to a few years, empirical statistics can be computed, even though the error bars are quite large.
This is not comfortable, but this is the best we can do.

Most of the figures included in this document have been computed with the time series for the MSCI index ``World Net Total Return USD Index'', Bloomberg ticker 'SPTR', and called ``Developed World Equity'' for short in this document.
This index represents the worldwide state of the stock market, summarizing many indexes.
For the bond market, the index ``Bloomberg Barclays US Agg Total Return'' is used, ticker 'LBUSTRUU', and called 'U.S. Aggregate Bonds' for short in this document.

The empirical samples used for most figures in this document start on 2000-01-31 and end on 2024-02-29 for all time series, giving a sample size of 24 years, with a good cross-section.
Clearly, the effective sample size is inconveniently small above one year, but this sample is characteristic of the recent market conditions.
In order to limit the number of figures, only computations done on this 2000-2024 sample are reported, together with simulations done on an equivalent 24 years period.

Some time series are available on longer samples, say starting in 1970, giving us 54 years of data, but with a much smaller cross-section.
Yet, the period 1970 to 1990 was characterized by a high inflation, and high interest rates, whereas the inflation is consistently low since 1990.
This difference distorts some statistics, in particular related to the long term mean returns and drifts, which become larger.
Otherwise, most statistics on distributions and lagged correlations give similar results on this longer sample.
Such consistency with a longer sample (but a smaller cross-section) give us some confidence about the estimation of long term statistical properties.

Underlying this discussion is a stationary hypothesis for the economy, clearly important when doing projection at the scale of decades.
Beside the changing level for the inflation and interest rates, no obvious changes are observed in the historical data.
This stationarity brings support to estimate the CMA over a few decades of historical data, and to the process used to simulated the possible market evolution.
Going beyond that would require to include in the model the inflation level, and its relation to interest rates and fixed income indexes.
This is clearly an interesting econometric project, which is left for further investigations.

\section{The drift}
\label{sec:drift}
\subsection{Overview}
\label{subsec:drift_overview}
As discussed in the introduction, the drift $\bm{\mu}(t)$ is an essential part of the process equation for long term simulations.
For a time horizon of 15 years and above, the random process is always in the drift dominated regime, hence the crucial importance of the drift for long term simulations.
The simplest model for the drift is a constant value, say as given by the CMA parameters
\begin{equation}
	\label{eq:drift_CMA}
	\bm{\mu}(t) = \bm{\mu}_\CMA.
\end{equation}
This base model is used almost universally for simulations, with the value for $\mu$ representing the long term grow of the economy (and the inflation) for the corresponding index.
For long-term Monte Carlo simulations, two modifications of this base constant model can be made.

First, the value for $\mu_\CMA$ is indeed a long term point forecast, a notoriously difficult number to evaluate, and which contains an error compared to the unknown realized value.
The past is certainly of some help, and the mean historical drift over the past few decades can be used (with a stationary hypothesis  to project the past values in the future).
Yet, large error bars should be assumed.
This error is quantitatively important, because the end value for the drift is $\DT\,\mu$, hence the error is also multiplied by $\DT$.
This issue is summarized in subsection~\ref{subsec:DU} below, and explored in depth for a simple model in appendix~\ref{annex:toy_model_drift_uncertainty}.
Assuming a normal distribution for the drift, its impact is investigated analytically on a simple normal random walk.
The analytic computation allows to understand the impact of the errors on the drift values, and the process is modified to include the uncertainty on the drift.
This modification is called the Drift Uncertainty (DU).
In the Monte Carlo simulations, the DU is included simply by altering the mean drift by a random term.
This term is defined and analyzed in subsection~\ref{subsec:DU}.
As an alternative, an econometric approach of the uncertainty on the CMA is presented in \citep{Ortec.UncertaintyInCMA}.

Lagged correlations for the returns and innovations point to positive correlations up to a few months, and negative correlations around a few years.
Such statistics are in agreement with short term ``trend following'' of traders, and with a ``mean reversion'' at longer time scales.
This behavior is embedded in the drift dynamics $\bm{\mu}(t)$ by lagged return terms, generating similar statistics in Monte Carlo simulations.
The most important term is the negative lagged correlations, which introduce a correcting effect after large drops or rises in the simulated prices, effectively ``stabilizing'' the long term process.
We have called this term \emph{Negative Return Correlation} (NRC).
The mechanism is similar to an AR (Auto Regressive) process, but for the present long term problem the drift equation has been modified in order to have simple terms acting at selected time scales.
This term is defined and analyzed in subsection~\ref{subsec:NRC}.


Consider now the standard deviation $\sigmaWealth$ of the terminal wealth of simulated paths at a given $\DT$, with the volatility for the paths given by $\sigma_\text{diffusion}$.
Let us emphasize that $\sigmaWealth$ measures the cross-sectional dispersion of the simulated values while $\sigma_\text{diffusion}$ measures the volatility in the time direction of the paths.
With a constant drift, the standard deviation is directly related to the volatility by
\begin{equation}
	\label{eq:sigma_path}
	\sigmaWealth(\DT) = \sqrt{\DT}\,\sigma_\text{diffusion}
\end{equation}
regardless of $\mu$.
Let us consider a stock index, with a typical annualized volatility of $\sigma_\text{diffusion} =$ 20\%.
At the scale of 16 years, these values lead to $\sigmaWealth$ = 80\%.
This is a very large uncertainty on the terminal values, leading to large probabilities for implausible large gains or losses (quantitatively depending also on $\mu$).
This feature is not realistic, since financial markets in the long term tend to correct themselves, with exuberant periods followed by crashes, and crashes followed by recoveries.
This is precisely the behavior introduced in the process by the negative return correlations.
Its impact is to lower $\sigmaWealth$ compared to $\sqrt{\DT}\,\sigma_\text{diffusion}$, modifying the relation \ref{eq:sigma_path}.
Then, the random uncertainty on the drift has the effect to increase the dispersion of the paths, effectively increasing $\sigmaWealth$ as $\DT^{3/2} = \DT^{1/2}\cdot\DT$ and altering the scaling in Eq.~\ref{eq:sigma_path}.
In short, the NRC term decreases $\sigmaWealth$ while the DU increases $\sigmaWealth$ proportionally to $\DT$.
Both terms in $\mu(t)$ modify $\sigmaWealth(\DT)$ and the parameters in the drift become important for the path dispersion, and for the control of the risk of an investment strategy.

\subsection{Drift Uncertainty (DU)}
\label{subsec:DU}
The base constant drift $\bm{\mu}_\CMA$ is a point forecast for the forthcoming realized drift, and as any forecast, it has an inherent uncertainty.
A better point of view is to replace the point forecast by a distributional forecast $p(\mu)$, acknowledging the uncertainty in $\mu$.
With this view, the CMA drift parameter is an estimate for the mean drift $\overline{\mu}$, and the related uncertainty parameter $\sigma_\mu$ should be estimated.
In a first step, the distribution for $\mu$ can be taken as a normal distribution.
A simple model along these lines is explored analytically in the appendix \ref{annex:toy_model_drift_uncertainty}.
In particular, it allows to relate the uncertainty $\sigma_\mu$ on $\mu$ to the volatility $\sigma_\text{diffusion}$ and to the length $\DT_\text{cal}$ of a (hypothetical) calibration sample used to estimate the historical drift $\hat{\mu}$.
The analytical result is that $\sigma_\mu$ is proportional to $\sigma_\text{diffusion}$ and inversely proportional to $\sqrt{\DT_\text{cal}}$, namely a good drift estimate is obtained with a small volatility and a long sample.

In Monte Carlo simulations, the uncertainty on the drift can be incorporated by adding a random component to the drift for each path using
\begin{equation}
	\label{eq:driftUncertainty}
	\mu_\text{DU}^{(k)} = \frac{\sigma_\CMA}{\sqrt{\DT_\text{cal}}}\, \epsilon^{(k)}
\end{equation}
where $\epsilon^{(k)} \sim \mathcal{N}(0,1)$ and $k$ indexes the paths.
In a multivariate context, the uncertainty $\epsilon^{(k)}$ is drawn independently for each time series.
The parameter $\DT_\text{cal}$ calibrates the importance of this term, and its quantitative impact on the standard deviation.

As shown in appendix \ref{annex:toy_model_drift_uncertainty}, the effect of the random drift is to increase the standard deviation of the prices at a given time horizon $\DT$ by
\begin{equation}
	\sigmaWealth(\DT) \simeq \sqrt{\DT}\, \sigma_\text{diffusion, 1y} \left(1 + \frac{1}{2}\,\frac{\DT}{\DT_\text{cal}}\right).
\end{equation}
where $\sigma_\text{diffusion, 1y}$ is the annualized volatility along the paths.
In particular, the correction term grows as $\DT$, namely the drift uncertainty makes the uncertainty on the terminal value to increase faster than $\sqrt{\DT}$ when $\DT > \DT_\text{cal}$.
The numerical simulations have been done with $\DT_\text{cal}$ = 25 years, leading to an increase of $\sigmaWealth$ by a factor 1.5 after 25 years.

\subsection{Negative Return Correlations (NRC)}
\label{subsec:NRC}
A discussed around Eq.~\ref{eq:sigma_path}, the purely diffusive volatility $\sigma_\text{diffusion}$ leads to large dispersions for the standard deviation of the price $\sigmaWealth$ at long horizons, mainly for stock indices.
In particular, some paths can end in the ``bankrupt'' area, with very small values.
Such occurrences are possible for individual stocks, but unlikely for indices constructed from the largest stocks on an exchange.
In order to have realistic long term distributions, a long-term mechanism to reduce the dispersion needs to be introduced.

This is done with a negative correlation between returns, introduced at a few selected time interval $\DT_k$.
The NRC drift component for a given asset is introduced with the following term
\begin{equation}
	\label{def:NRC_equation}
	\mu_\text{NRC}(t) = \sum_k \gamma_k \,\frac{\dt}{\DT_k}\, \left(\frac{p(t)}{D^{-1}(\DT_k)\, p(t-\DT_k) } - 1 \right).
\end{equation}
The discount factor $D(\DT)$ is
\begin{equation}
	\label{def:discountFactor}
	D(\DT) = (1 + \mu_\CMA)^{-\DT/\dt},
\end{equation}
and $D^{-1}(\DT)$ moves forward the price by $\DT$.
In \eqref{def:discountFactor}, $\mu_\CMA$ is the drift at scale $\dt$ and plays the role of the interest rate.
Since $\DT_k$ can be of the order of years, it is important for the historical return evaluation to properly move to $t$ the price $p(t-\DT_k)$.
In Eq.~\ref{def:NRC_equation}, the term in parentheses is the return at scale $\DT_k$, computed at $t$.
The term $\dt/\DT_k$ is a convenient scaling factor so that all historical returns are scaled to $\dt$, regardless of $\DT_k$.
The coefficient $\gamma_k$ fixes the magnitude of the return correlation at the time scale $\DT_k$.
In order to obtain the actual history dependent drift, the NRC drift component is added to the base constant drift, and possibly a DU component is also added.

In Eq.~\ref{def:NRC_equation}, if $\gamma$ is positive, the past returns are amplified, while a negative $\gamma$ dampens the price fluctuations by inducing a correction in the opposite direction.
The coefficients $\gamma_k$ have been adjusted so that the process match the empirical statistics for the lag-one correlations for the returns and innovations, see the figure \ref{fig:LaggedCorrelation_returnHist_Vs_returnReal} for such an investigation.
Two time horizons $\DT_k$ already provide for a satisfactory agreement with empirical lagged correlations for the returns, while 3 time horizons do not improve significantly.
The parameter $\gamma_k$ can in principle be adjusted for each index, we found that it is sufficient to have values depending only on the asset class.
This is consistent with the view that all the stock indexes behave statistically in a similar way, but stock and bond indexes could have a different dynamics.

In finance, many diffusion models include a mean reverting component, mainly to describe interest rate dynamics or stochastic volatility dynamics.
Generically, they are known as Orstein-Uhlenbeck (OU) processes, or as CIR processes for interest rates, with analytical solutions for the simplest specifications.
The drift equation in a OU process can be adapted to our long-term simulation horizon.
An OU diffusion acts similarly as the NRC term by limiting the diffusion, but over the full drifted path for a OU term while only at the scale $\DT_k$ for the NRC.
The OU term is stronger, leading to a constant asymptotic standard deviation, namely $\sigma_\text{value}(\DT)$ is not growing with $\sqrt{\DT}$ but converges to a constant for large $\DT$.
This limited diffusion behavior is realistic for interest rates and implied volatilities that are fundamentally bounded, but not for stock indexes which can grow without limitation.
Hence our focus on the NRC term.

\subsection{Empirical validation: NRC}
\label{subsec:empirical_validation_NRC}
\begin{figure}[ht]
	\centering
	\begin{subfigure}{0.8\textwidth}
		\includegraphics[width=\textwidth]{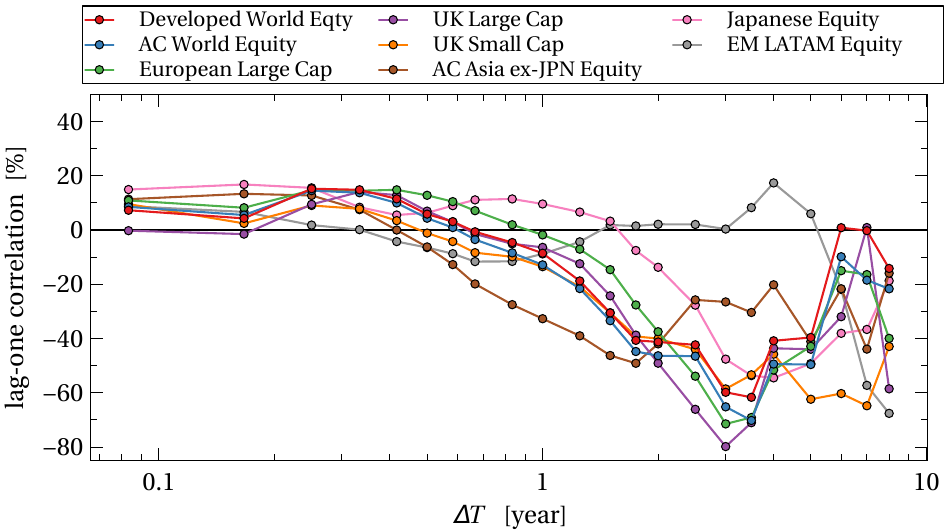}
	\end{subfigure}\vspace*{2ex}
	\begin{subfigure}{0.8\textwidth}
		\centering
		\includegraphics[width=\textwidth]{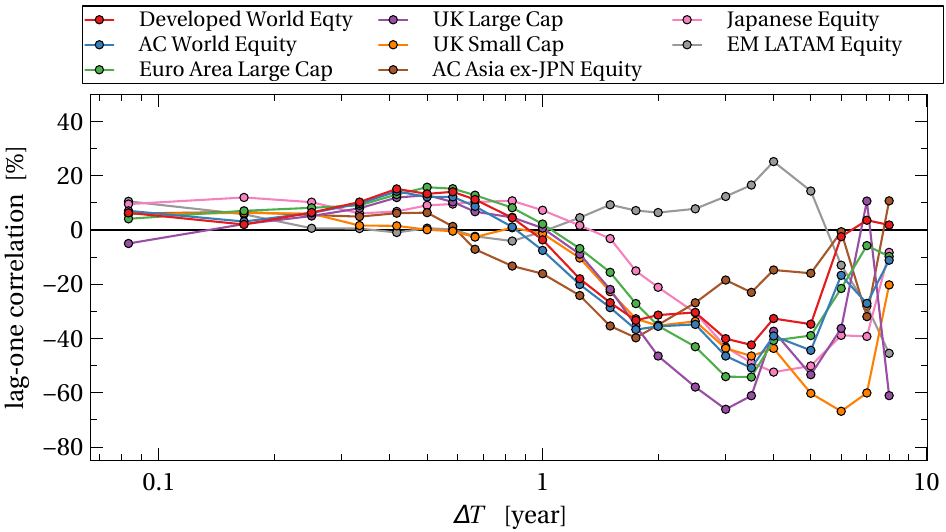}
	\end{subfigure}
	\caption{Lag-one correlations for the returns (top) and innovations (bottom), for equity indices.}
	\label{fig:lagCorr_scaling_equities}
\end{figure}

\begin{figure}[ht]
	\centering
	\begin{subfigure}{0.8\textwidth}
		\includegraphics[width=\textwidth]{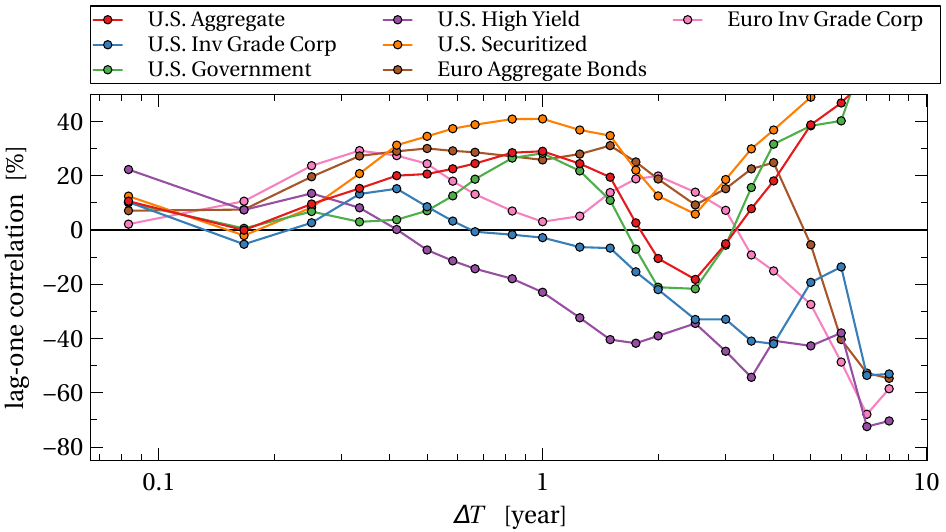}
	\end{subfigure}\vspace*{2ex}
	\begin{subfigure}{0.8\textwidth}
		\includegraphics[width=\textwidth]{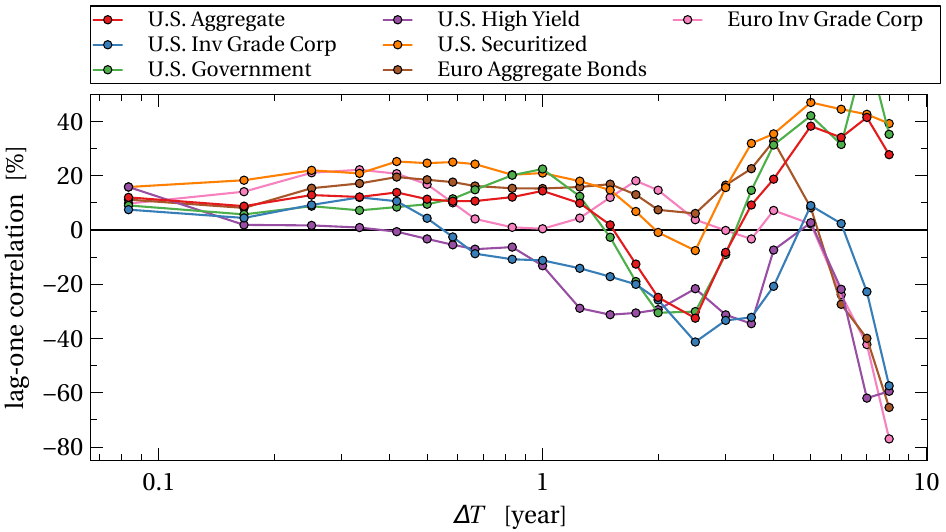}
	\end{subfigure}
	\caption{Lag-one correlations for the returns (top) and innovations (bottom), for fixed income indices.}
	\label{fig:lagCorr_scaling_fixedIncomes}
\end{figure}
A drift term depending on the past values introduces lagged correlations between returns and between innovations (for $\DT > \dt$).
This is most efficiently measured by using the ``lag-one'' correlation for the returns and the innovations, as described in Sec.~\ref{subsec:lagOneCorrelation}.
Figures \ref{fig:lagCorr_scaling_equities} and \ref{fig:lagCorr_scaling_fixedIncomes} display the lag-one correlations respectively for a set of stock indices and bond indices, computed with a sample from 2000-01-31 to 2024-02-29.
The similarity between all the curves points to a universal behavior (within each asset class), that can be reproduced with one set of parameters (per asset class) for the NRC term.
The same computation for a few time series starting in 1970 gives a similar figure, as well as a larger sample on a shorter time span.
Hence, these empirical correlations seem fairly robust against the choice of time series and sample periods (the fixed income indices being the most sensitive to the start and end dates).
Yet, the substantial correlations between markets introduce an unquantified dependency between the lag-one correlations, making dependent the various curves in these figures.
The weakest dependency is between equities and bonds, and the similarities between Fig~\ref{fig:lagCorr_scaling_equities} and \ref{fig:lagCorr_scaling_fixedIncomes} point to a similar behavior.

On the empirical figure for equity indexes, positive lagged correlations are observed at a scale of one to a few months.
Likely, these positive correlations result from trend following market participants.
At the scale of 1 year to a few years, negative lagged correlations are observed, with values of the order of -30\% to -60\% for stock indices.
These negative correlations indicate medium term market corrections, with moves in one direction followed by a move in the opposite direction.

The short term interest rates have positive lagged correlations, say up to 1 year, likely due to the decision pattern of the central banks: their decisions on the overnight rates are to increase or decrease by small increments but many times and over a long period.
In turn, these decisions propagate toward longer maturities along the yield curves.
This decision pattern induces clear lagged correlations, mainly for the short term bonds, but also for longer maturities.
Beside, the main bond markets are US and Europe, and these central banks have correlated behaviors.
These two effects make difficult the construction of a set of independent bond indexes.
The indexes used for the figure \ref{fig:lagCorr_scaling_fixedIncomes} are US and European specific indexes, mitigating somehow these correlations.
At the scale of 1 year to a few years, negative lagged correlations are observed, with values of the order of  10 to -30\% for bond indices.

\begin{figure}[ht]
	\figurePannelFlat{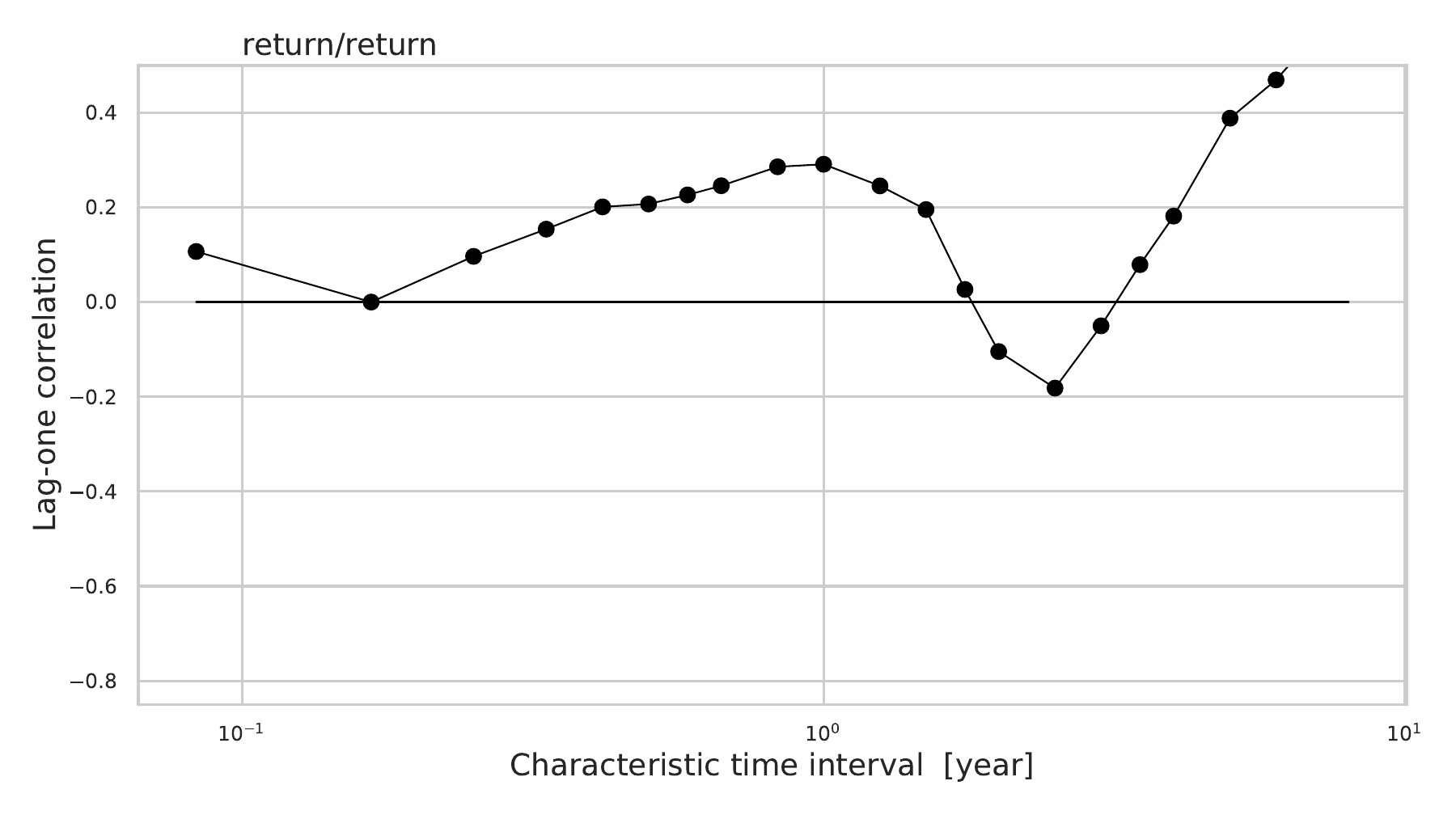}
	\caption{
		The lag-one correlations between the returns (historical and realized), versus the time horizon $\DT$ on the x-axis, for empirical data and various processes.}
	\label{fig:LaggedCorrelation_returnHist_Vs_returnReal}
\end{figure}

Based on the empirical correlations, 2 terms have been added in the drift according to Eq.~\ref{def:NRC_equation}, at scale of 6 months with a positive coefficient and at scales 40 months with negative coefficients.
The resulting lag-one correlations are plotted in \ref{fig:LaggedCorrelation_returnHist_Vs_returnReal}, which reproduces the main characteristics of figures \ref{fig:lagCorr_scaling_equities} and \ref{fig:lagCorr_scaling_fixedIncomes}.
For the 2 models without NRC (line b and c on the figure), notice the downward tendency for large $\DT$, whereas these models have a zero lagged correlation for all $\DT$.
This tendency is due to a small sample bias for this statistical estimator, which is hidden by the lagged correlations in line a and d.

\FloatBarrier
\section{The covariance}
\label{sec:covariance}
\subsection{Overview}
\label{subsec:convariance_introduction}
The covariance matrix is the multivariate extension of the squared volatility, and it controls several aspect of the process.
First, it fixes the volatility of each indexes, and the correlations between them.
This can be achieved by a constant covariance, with $n(n+1)/2$ parameters.
A constant covariance term also fixes the long term properties of more sophisticated model.
Second, a dynamic component can be added, in the spirit of the GARCH(1, 1) process for the univariate case.
The vast family of ARCH-like processes provides for many analytical structures that can capture features of empirical time series with various level of details.
At the core, these processes can model the quiet periods, agitated periods, and crises that are observed in the economy and in the financial time series.

A vast literature exists on the multivariate extension of ARCH like models, in general leading to very complex model for increasing size $n$.
The Multivariate Generalized Autoregressive Conditional Heteroskedasticity (MGARCH) model, along with its extensions, could be used for multivariate portfolio simulation, see e.g. \citep{BauwensLaurentRombouts.2006} for a review.
Yet, many of these models suffer from a significant limitation because the number of parameters increases rapidly with the number of assets.
For instance in the BEKK(1,1,1) model (see \citep{BEKK}), there are $n(5n+1)/2$ parameters, resulting in 24 parameters for only three assets.
Such models are clearly unsuitable for a large universe, due to the quadratic grow of the number of parameters that must be estimated.

In order to construct a simple model regardless of the universe size, we are using a convex combination between a constant covariance and a dynamical part built from a Long Memory ARCH (LMARCH) process.
The LMARCH component captures efficiently the dynamics of the volatility, and importantly with the same few parameters for all time series.
The convex combination between the static and dynamic estimations is controlled by 1 parameter denoted by $w_\infty$, building an affine model for the covariance (as a function of the squared returns).
This model is minimal, yet describes correctly the key aspects of the multivariate volatility dynamics.

The static and dynamic components of the covariance matrix are introduced in the next two subsections, followed by a comparison between the lagged correlation of the volatility for empirical data and for numerical simulations.

\subsection{Constant Covariance}
\label{subsec:convariance_constantCovariance}
The simplest possible model for the covariance is to take a constant matrix, with the values given by the CMA parameters.
The covariance is conveniently decomposed into volatility and correlation using
\begin{align}
	\bm{\Sigma}(t) =  \bm{\Sigma}_\CMA =  \bm{\sigma}_\CMA\cdot \bm{\rho}_\CMA \cdot\bm{\sigma}_\CMA
	\label{eq:sigma_from_rho}
\end{align}
where $\bm{\rho_\CMA}$ is the correlation matrix and $\bm{\sigma}_\CMA$ the diagonal matrix of volatilities.
This model has $n$ parameters for the volatilities and $n(n-1)/2$ parameters for the correlations, leading to a total of $n(n+1)/2$ parameters.
These parameters are set in the CMA (Capital Market Assumptions).
For Monte Carlo simulations, the constant covariance has the advantage that its square root can be computed once, leading to fast simulations.

In general, the volatilities provided in the CMA are at an annual scale.
They must be scaled at $\dt$ using a factor $\sqrt(\dt/\oneYear)$.

\subsection{LMARCH Covariance}
\label{subsec:lmarchProcess}
For univariate time series, the LMARCH process proved to be a very convenient workhorse for several financial applications, see e.g. \citep{Zumbach.book} and the reference therein.
The long memory embedded in the weights captures correctly the slow decay of the correlations for the volatility.
Another point of view is that the clustering of the volatility occurs at time scales ranging from days to years, and a processes with one time scale like I-GARCH(1) or GARCH(1,1) are unable to reproduce such behaviors which requires a multi-scale model.
Another very interesting property of the LMARCH model is its very sparse parametrization, and the ability to describe well most time series with the same small set of parameters.
Essentially, the memory kernel for the volatility depends on 1 parameter fixing the decay of the weights, while the other parameters are essentially cut-offs that are less important.
For these reasons, this process can be used for many applications (risk valuation, portfolio covariance estimate, option pricing), without the necessity to optimize parameters.

As given by Eq.~\ref{def:varianceForecast}, the univariate linear LMARCH process evaluates the covariance with a sum of past squared returns, with weights decaying gradually as the lag increases.
Because of its simple quadratic structure, it generalizes naturally in the multivariate case
\begin{align}
	\label{eq:lmarch_linear}
	&	\Sigma_\text{LM-ARCH, linear}^{\alpha, \beta}(t; \dt) = \\
	&	\hspace{2em} \sum_{0 \leq l < l_\text{max}}  w(l, \dt) ~\left(r_\alpha(t - l\,\dt; \dt) - \mu_\alpha\right)
	\cdot \left( r_\beta(t - l\,\dt; \dt) - \mu_\beta\right). \nonumber
\end{align}
where $\alpha$ and $\beta$ index the time series.
Notice that the weight function $ w(l, \dt)$ does not have a dependency on $\alpha$ and $\beta$, namely the weights are independent of the time series.
This simple extension was used in \citep{Zumbach.largeCovarianceMatrices} to investigate the empirical properties of large covariance matrices.

A process with the covariance defined by Eq.~\ref{eq:lmarch_linear} is purely auto-regressive, similar to an I-GARCH process for the univariate case.
For Monte Carlo simulations, the mean volatility is not defined by the covariance, and such processes are unstable (more precisely the asymptotic probability distribution is singular, see \citep{Nelson.1990, Corradi.2000} for an analysis using the I-GARCH(1) model).
A mean covariance must be added in order to stabilize the process for long term simulations, leading to an affine model for the covariance (affine in term of the squared returns and covariance).
For the multivariate case, this leads to the covariance
\begin{align}
	\label{eq:lmarch_affine}
	\bm{\Sigma}(t;\dt) = \bm{\Sigma}_\text{LM-ARCH, affine}(t) = \wInfty \cdot \bm{\Sigma}_\CMA + (1 - \wInfty) \cdot \bm{\Sigma}_\text{LM-ARCH, linear}(t; \dt)
\end{align}
with $0 \leq \wInfty \leq 1$.
The parameter $\wInfty$ controls the balance between the long term covariance and the auto-regressive covariance.
Large values for $\wInfty$ lead to a model close to a constant covariance, with small volatility clustering, narrow distributions for the volatility, and small lagged correlations for the volatility, while small values for $\wInfty$ leads to larger excursion of volatility, in line with typical market behavior.

For the Monte Carlo simulations, the parameter $\wInfty$ has been adjusted so that the probability distributions  and the lagged correlations of the volatility match at best between empirical and simulated time series.
These statistics point to low value for $\wInfty$.
Yet, with long term simulations, a small value for $\wInfty$ can lead to small values for the simulated prices.
In practice, an index value $p(t)$ smaller than 1/100 of the initial value $p(0)$ is considered as unrealistic (at least for the indexes used in this study).
A similar argument can be made for unrealistic large values.
This condition induces a lower bound on $\wInfty$, given by the probability of too small prices to be negligible.

This trade-off is also influenced by the model used for the drift, either constant or with a NRC term.
Negative return correlations stabilize the long term process, reducing the width for the terminal wealth's distribution, hence diminishing the probability for unrealistic low or high values.
With a NRC term in the drift, a lower value for $\wInfty$ can be used, leading to a better agreement with empirical statistics.

\subsection{Empirical validation: volatility}
\label{subsec:vol-on-vol_empirical}

\begin{figure}[ht]
	\centering
	\begin{subfigure}{0.8\textwidth}
		\includegraphics[width=\textwidth]{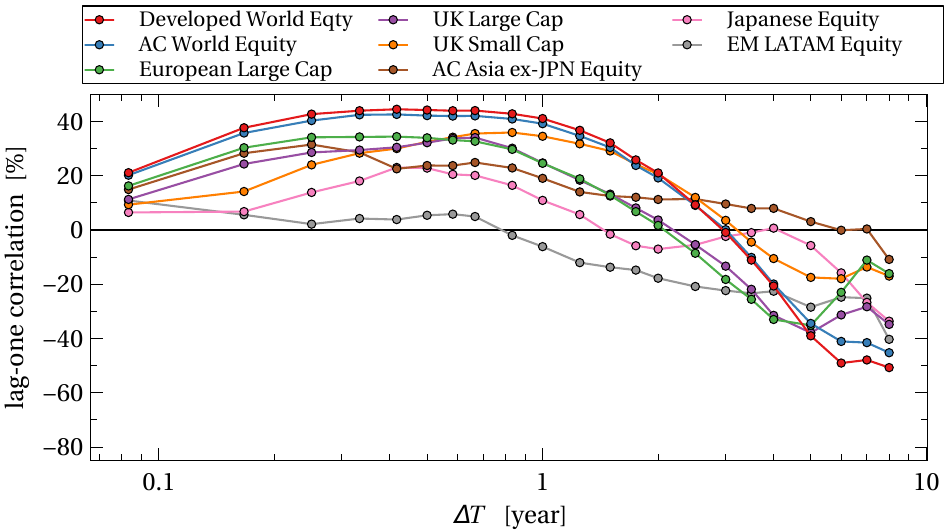}
	\end{subfigure}\vspace*{2ex}
	\begin{subfigure}{0.8\textwidth}
		\includegraphics[width=\textwidth]{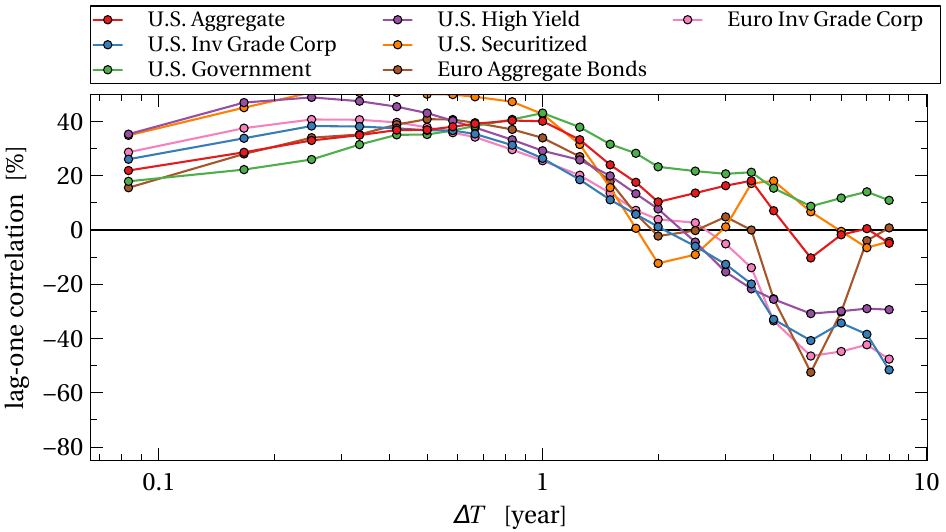}
	\end{subfigure}
	\caption{Lag-one correlations for the volatilities for equity indexes (top) and fixed income indices (bottom).}
	\label{fig:volOnVol_lagCorr}
\end{figure}
The volatility dynamics is measured by the lagged correlations for the volatility, with their definitions given in Sec.~\ref{sec:quantities}.
The graph~\ref{fig:volOnVol_lagCorr} shows the empirical lag-one correlations between the historical LMARCH volatility and the realized RMA volatilities at increasing time scale $\DT$.
The parameters $\DT$ is on the horizontal axis, and the lagged correlations are evaluated for a selection of equity and fixed income indices.
Essentially, these graphs measure the dependency from the past of the volatility at increasing time horizons $\DT$, from 1 month to 8 years.
Overall, the correlation is in the 20 to 40\% range for $\DT$ between a few months to 1 year.
This is quantitatively the largest dynamical effect of financial time series, showing the importance of the volatility clustering.

\begin{figure}[ht]
	\figurePannelFlat{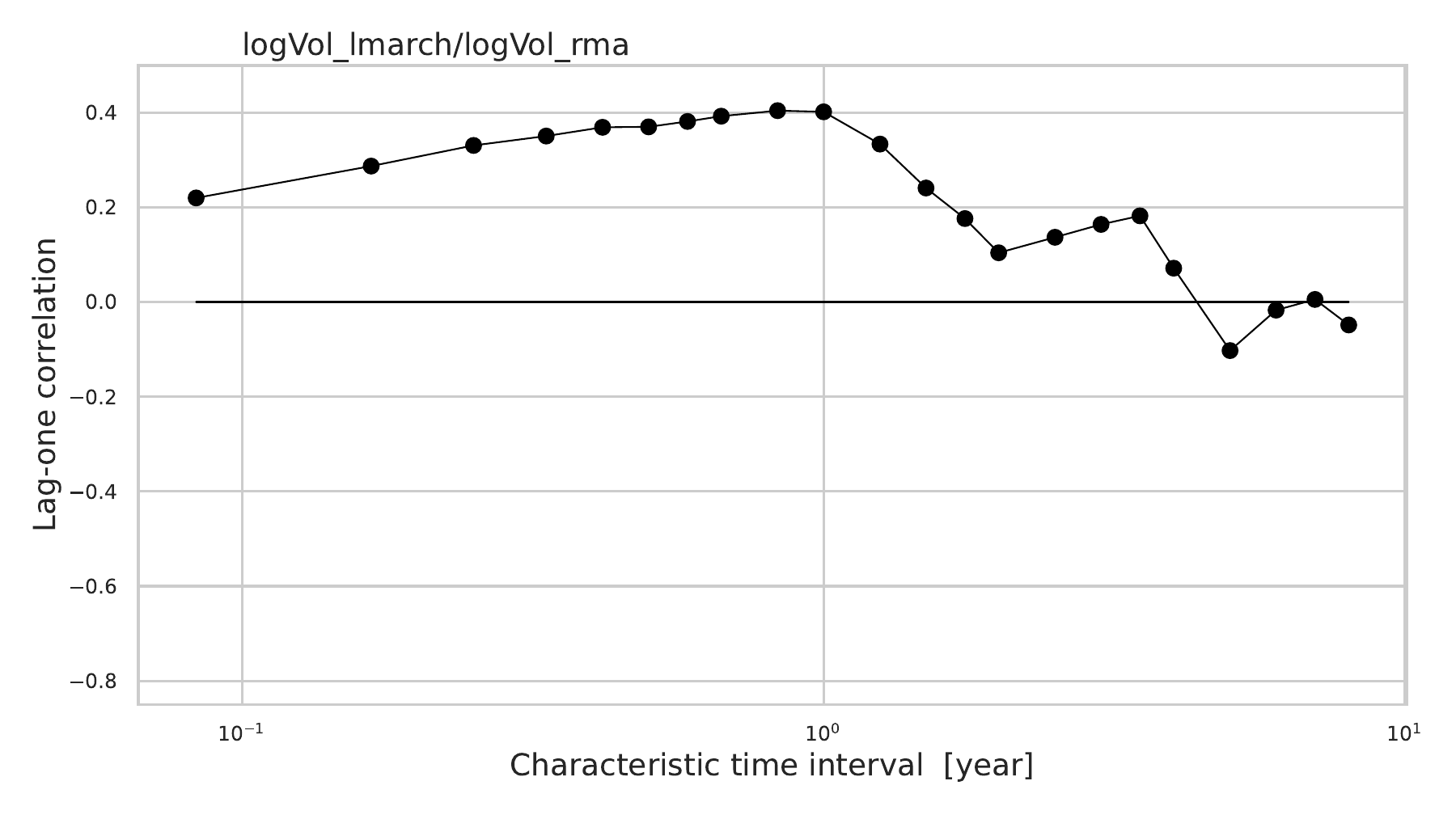}
	\caption{Lag-one correlations for the volatility. Empirical (top) and for 4 processes.}
	\label{fig:LaggedCorrelation_logVol_lmarchHist_Vs_logVol_rmaReal}
\end{figure}
The graph~\ref{fig:LaggedCorrelation_logVol_lmarchHist_Vs_logVol_rmaReal} shows the ``lag-one'' correlation of the volatility for some processes.
The graphs on line a are based on empirical time series, with the indexes given in the caption.
The graphs in line b use a process with constant covariance.
Since the volatility dynamics is absent for this process, this correlation is zero.
Notice the negative biases due to the small effective sample at increasing $\DT$.
The graphs on line c are computed with no NRC term, a LMARCH process (with $w_\infty = 0.55$), and a non-central Student distribution,
while the graphs on line d are computed with a NRC term, a LMARCH process (with $w_\infty = 0.40$), and a non-central Student distribution.
A process with no NRC and $w_\infty = 0.40$ is not shown: it has a finite probability to end in the ``bankrupt'' domain with too small values, which is not realistic.
Essentially, the NRC term is acting to lower the terminal wealth variance.
The qualitative behavior reproduces well the empirical correlations, albeit quantitatively on the small side.
More sophisticated models would be needed to reach a better quantitative agreement.

\begin{figure}[ht]
	\figurePannelFlat{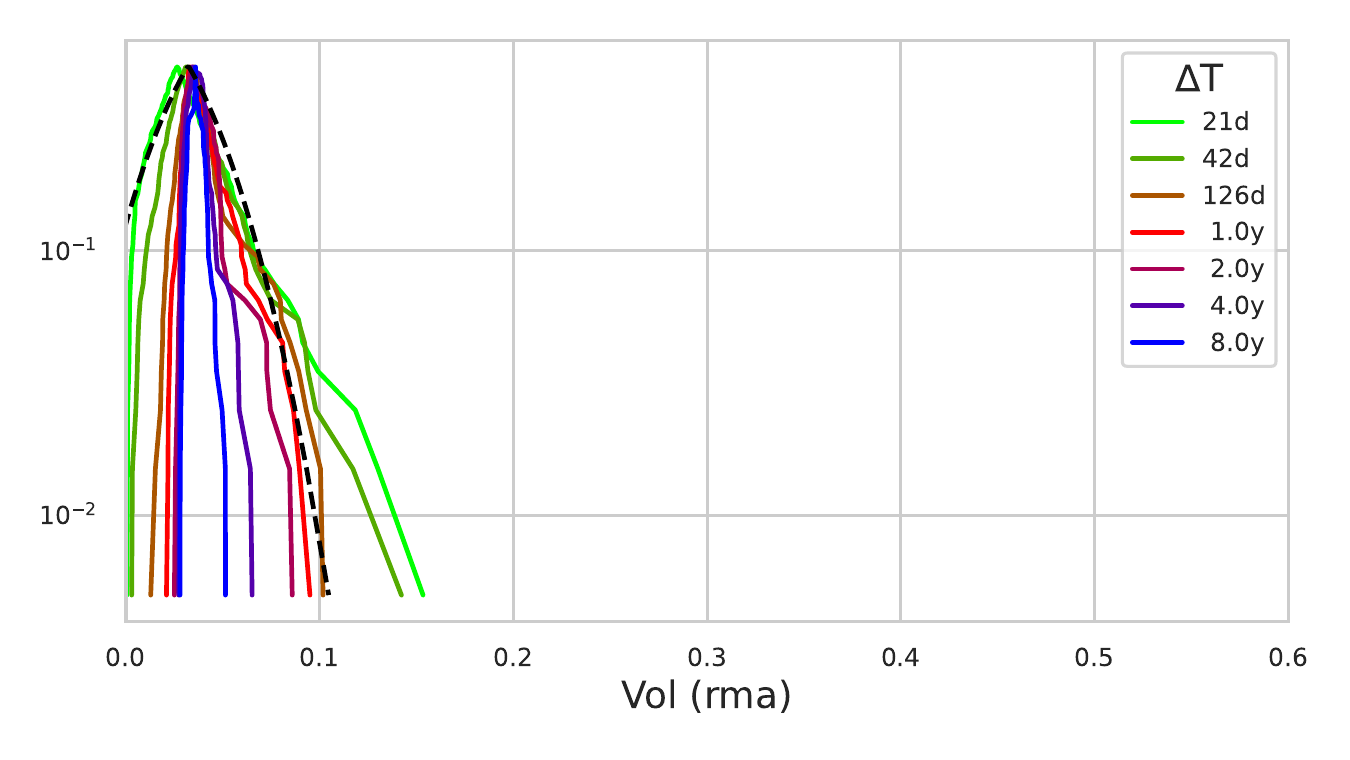}
	\caption{The folded-cdf for the volatility (measured with equal weights in the interval $\DT$), at increasing time horizon $\DT$.}
	\label{fig:FoldedCdf_vol_lma}
\end{figure}
The distribution for the volatility is important since depending on its dynamics, and controlling partly both tails of the return distributions.
Figure \ref{fig:FoldedCdf_vol_lma} shows the folded-cdf for the RMA volatility over the period $\DT$, and annualized.
The figure compares the empirical distributions for the same processes.
Clearly, a process with constant covariance cannot reproduce the broader distributions of the empirical volatility.
Another default of the constant covariance is the fast convergence to the long term volatility, since the width of the distribution is depending only on the number of squared returns in the evaluation of the volatility (i.e. decreasing as $\sqrt{\text{1m}/\DT}$ for monthly returns).
The bottom figure shows the good agreement between the LMARCH process and the empirical cdf, emphasizing the impact of the volatility dynamics on the distribution.
As a caveat, the CMA used for the simulations have been computed on the same empirical time period, hence the very good agreement for the median between empirical and simulated graphs.
In a true out-of-sample simulation, the agreement would not be necessarily of this quality.

\FloatBarrier
\section{Distribution for the returns and innovations}
\label{sec:randomReturns}
\subsection{Random generators and non-central Student distribution}
\label{subsec:randomGenerator}
The last important part of the process that needs to be specified is the distribution for the innovations.
The standard choice is to use a multivariate normal $p_{\bm{\epsilon}} = N(\bm{0}, \bm{\id})$, leading to the distribution for the returns $p_{\bm{r}} = N(\bm{\mu}, \bm{\Sigma})$, and this distribution decays exponentially fast.
As shown by several empirical studies, financial markets experience more brutal events compared to the exponential decay of the normal distribution.
This is remedied by using instead a Student distribution, which decays as a power law for large innovations.
For both normal and Student distributions, the multivariate extensions are well known and can be found in many textbooks, see for example \citep{Embrechts.2015}.

In the univariate case, the empirical distributions from the returns can be directly obtained from the prices.
Removing the heteroskedasticity is interesting since leading to stationary time series, with better convergence properties for the cdf.
The innovations can be obtained by using Eq.~\ref{def:innovations_dt} at scale $\DT$, computing the return and volatility on historical data, and finally computing the time series of realized innovations and their distributions.
A normal distribution clearly lacks heavy tails, while a Student distribution matches decently both empirical distributions, with a tail index of the order of 3 to 10 at the monthly scale.
Yet, the empirical distributions for the returns and innovations are asymmetric, with larger moves on the down side than on the upside, particularly for stock indexes but also for bond indexes.
A Student distribution is symmetric, therefore cannot reproduce this market asymmetry.

A non-central Student distribution provides for a simple solution, where the asymmetry is controlled by one new parameter $\gamma$, and with a simple algorithm to generate random draws.
The distribution and related random generator is presented for example in \citep{Embrechts.2015}, or  with slightly different specification in \citep{wikipedia_nonCentralT}.
Remain to have a non-central Student random generator for the multivariate case.
Such an algorithm is given in \citep{Embrechts.2015} when presenting the ``normal mean-variance mixture'' algorithms, depending on the parameters $\bm{\mu}$, $\bm{\Sigma}$, and a vector $\bm{\gamma}$ specifying the asymmetry.
For $\bm{\gamma} = 0$, the multivariate non-central Student distribution reduces to the usual Student distribution.

\begin{figure}[ht]
	\centering
	\includegraphics[width=0.65\linewidth]{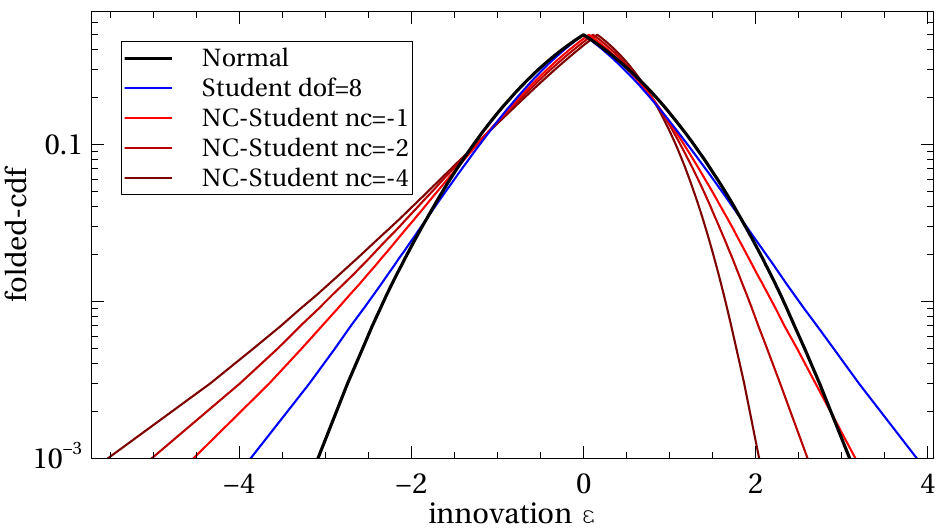}
	\caption{The folded-cdf f-cdf($\epsilon$) for different distributions for the innovation $\epsilon$ in the univariate case.
		All Student have 8 degree of freedoms}
	\label{fig:student_distribution}
\end{figure}

Yet, the equations provided in \citep{Embrechts.2015} are mixing the parameters, in particular the mean and covariance of the generated vectors is not equal to $\bm{\mu}$ and $\bm{\Sigma}$, respectively, but have contributions in $\bm{\gamma}$.
As provided, this parameter's mixing makes the algorithm difficult to use in an application.
Fortunately, the equations can be modified so that the means and the covariance matrix of the generated random vectors are given by the corresponding parameters.
This part is a bit technical, and is presented in Appendix~\ref{appendix:non-central-student}.
The key result is that multi-variate random deviates $\bm{\epsilon}$ can be generated with zero mean, unit variance, fat tails, and the asymmetry specified by $\bm{\gamma}$.
Figure \ref{fig:student_distribution} shows the main distributions in the univariate case, namely normal, Student and non-central Student for increasing non-centrality parameter.

The vector of random returns $\bm{r}$ is obtained with
\begin{align}
	\label{eq:random_non_central_student_return}
	\bm{r} & = \bm{\mu} + \bm{A} \cdot \bm{\epsilon}
\end{align}
where $\bm{A}$ is a square root of $\bm{\Sigma}$.
Using the expectation and variance of $\bm{\epsilon}$, simple computations leads to
\begin{subequations}
	\label{eq:mean_and_variance_for_r}
	\begin{align}
		\E{\bm{r}} & = \bm{\mu} \\
		\var{\bm{r}} & = \bm{\Sigma}.
	\end{align}
\end{subequations}

The parameters values for $\bm{\gamma}$ have been adjusted on empirical data.
A cross-sectional study shows that it is sufficient to have one value per broad sector, like stock indexes or fixed incomes indexes.
These studies show that the asymmetry is always on the negative side (larger down moves than up moves), and with a slightly larger asymmetry for stock indexes than for fixed income indexes.

\newpage
\subsection{Empirical validation: innovation distributions}
\label{subsec:innovation_distributions}
\begin{figure}[ht]
	\figurePannelFlat{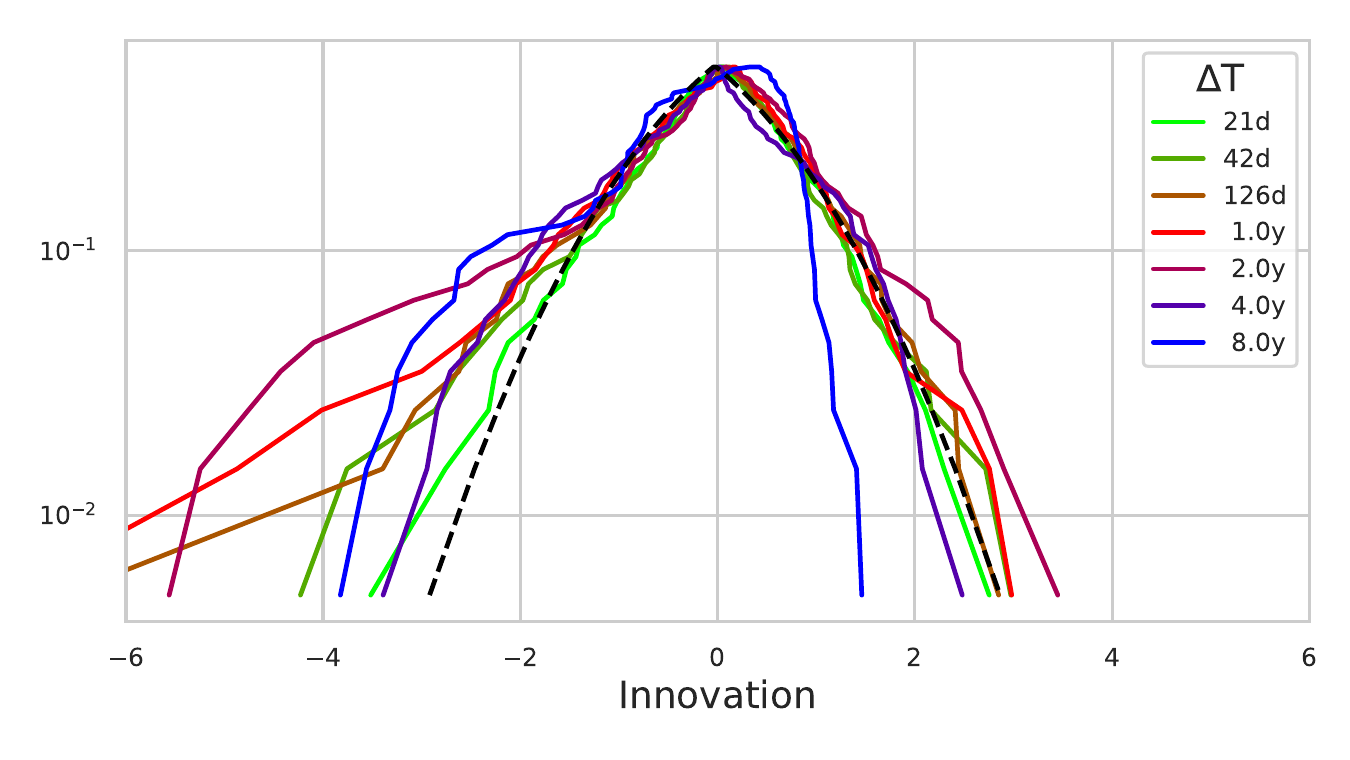}	\centering
	\caption{The folded-cdf f-cdf($\epsilon(\DT)$) for the innovations $\epsilon(\DT)$ at increasing time horizon $\DT$.}
	\label{fig:FoldedCdf_innovation}
\end{figure}
The empirical and simulated distributions of the innovations for the standard visualization set is shown in Fig.~\ref{fig:FoldedCdf_innovation}.
All the figures are generated from whole price paths, historical market prices (line a) or simulated Monte Carlo prices (line b, c, d), on which the formula \ref{def:innovations_dt} is used to compute the innovations $\epsilon(\DT)$ at scale $\DT$.
On the empirical graphs (line a), the deep tails asymmetry is clearly visible, and is due to strong down moves (crashes) occurring after a period of low volatility (e.g. Covid crisis, Ukraine war).
The asymmetry is particularly strong at the scales of 2 to 6 months, with quite large negative innovations.
Such large persistent asymmetry cannot be reproduced with the current process. 
A term braking the $r \rightarrow -r$ symmetry should be introduced, like a leverage effect term in the volatility (the volatility becomes larger following a price drop).
Introducing such a term in a multivariate context will complicate significantly the model, and we decided not to pursue this venue.

\subsection{Empirical validation: return distributions}
\label{subsec:return_distributions}
\begin{figure}[ht]
	\figurePannelFlat{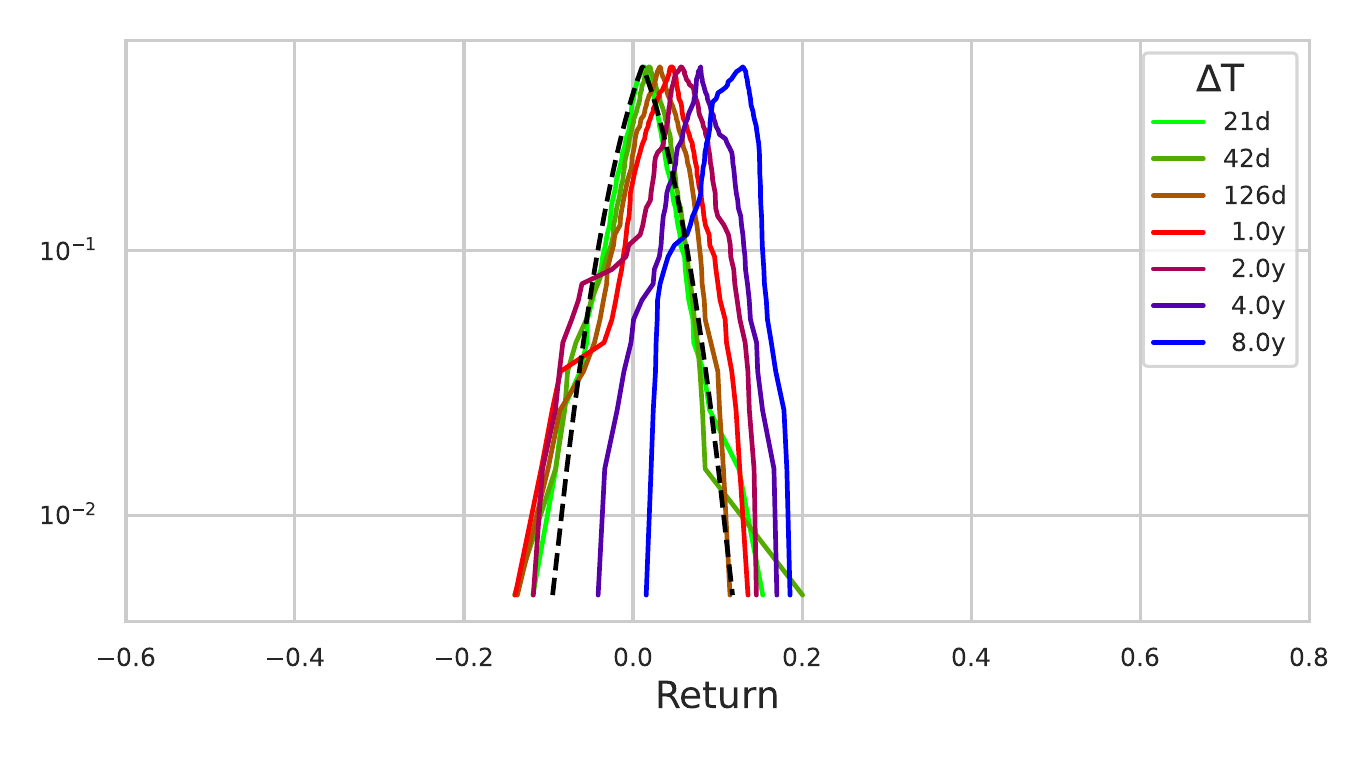}
	\caption{The folded-cdf f-cdf($r(\DT)$) for the annualized return $r(\DT)$ at increasing time horizon $\DT$.}
	\label{fig:FoldedCdf_return}
\end{figure}
The same plot but for the returns $r(\DT)$ is shown in Fig.~\ref{fig:FoldedCdf_return}.
For this figure, the returns are computed with a logarithmic return formula $r(t, \DT) = \log(p(t)/p(t-\DT))$, because both prices play a symmetric role, leading to more symmetric distributions.
The returns at scale $\DT$ are annualized with a factor $\sqrt{1y/\DT}$, resulting in comparable distribution shapes.
The location parameter (mean or median) should be annualized with a factor $1y/\DT$, hence a remaining scaling factor  $\sqrt{1y/\DT}$ for the median on these distributions.
This simple scaling difference explains the shift to higher values for the median at increasing $\DT$ (i.e. the point at cdf = 1/2).
Overall, this figure shows the important similarity in the shape of the distributions when increasing $\DT$.
The black line is the cdf for a normal distribution with the same location and size as the 1 month returns.
The asymmetry and fat-tails of the distributions are clearly visible.
Interestingly, the distributions at 3 and 6 months show even larger tail and asymmetry, a feature observed in most empirical time series.

The corresponding figures for Monte Carlo simulations are displayed on the following lines, with an overall fair agreement with the empirical distributions.
For these processes, the aggregation necessarily reduces the tails and asymmetry, and more sophisticated processes are required to model the respective increases at a few months scale.
The agreement for the width of the distribution at 1 month results from using the CMA computed on historical data, hence is not surprising.
The interesting feature is the reducing width when increasing $\DT$, which is similar for the empirical and simulated time series.
This shows that the LMARCH process captures correctly the time aggregation of the returns.

\FloatBarrier
\section{Processes comparison}
\label{sec:processesComparison}
The main new features of the processes are the ARCH volatility, together with the NRC and DU terms that modify the link between the standard deviation of the simulated values from the volatility in the time direction.
Another notable feature is the dependency from the past at the simulation start introduced by the LMARCH volatility and the NRC term.
Different processes are compared in the following subsections, using Monte Carlo simulations with 50,000 paths, and a simulation length of 20 years with monthly steps.

The start date is May 31, 2020, namely when the Covid crisis was in full swing.
At this date, the market was lower and with a high volatility, but already recovering from the sharp plunge in February.
This particular date was selected in order to show clearly the impact of the initial conditions, corresponding to a major crisis.
Outside of this particular period, the market is more ``typical'', with less dependency on the initial conditions.

Five processes have been selected for the comparison, with combinations of the followings.
The base drift is constant, and a NRC and/or DU terms can be added.
The volatility can be constant or with a LMARCH covariance.
The innovations can be normal, or an asymmetric Student.
The parameters for the various components are chosen so as to match overall the corresponding statistical properties, as described in the previous sections.
For the DU term, the calibration time span $\DT_\text{cal}$ in Eq.~\ref{eq:driftUncertainty} is taken as 25 years.
Five combinations of the above have been selected in order to show the salient features introduced by each term.
The base reference process is the combination (constant drift, constant volatility, normal innovations), corresponding to the usual (multiplicative) normal random walk.

One asset is used for the simulation, with the CMA corresponding to the developed world equity.
Because only one asset is present, there is no rebalancing or strategy.
Notice that realistic simulations involve a portfolio together with a strategy (periodic rebalancing, cash-in or cash-out).
In this setting, the strategy can have a further impact on the wealth, depending on the process used for the simulations.
For example, a cash-out during a crisis will have a large impact, and a LMARCH volatility or an asymmetric Student innovation can create larger price moves.
For this reason, a realistic process must be used in actual simulations, but a study of portfolios and strategies is beyond the present scope.

\subsection{Drift for the wealth}
\begin{figure}[ht]
	\centering
	\includegraphics[width=0.8\figureWidth]{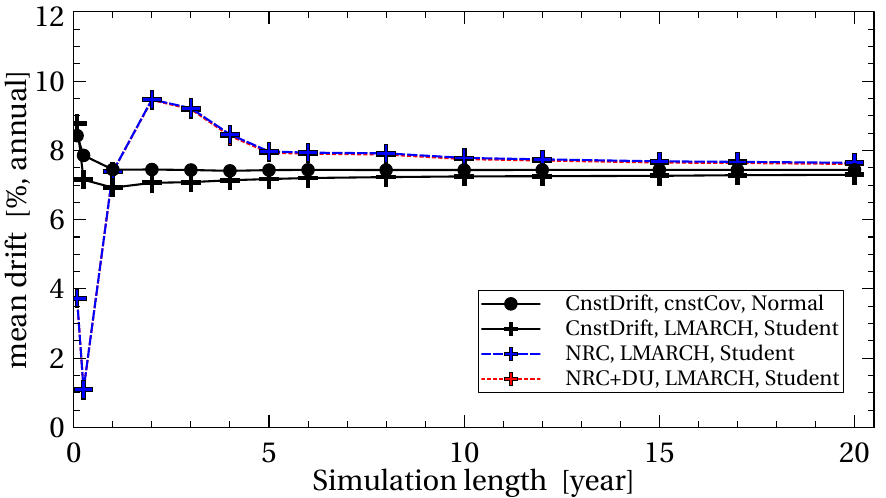}
	\caption{The mean drift for different processes, as function of the simulation length.}
	\label{fig:wealthDrift}
\end{figure}
The Fig.~\ref{fig:wealthDrift} shows the simulated expected drift for the selected processes.
The mean wealth value after the simulation time $\DT$ is annualized, so as to remove the leading $\DT$ scaling.
The processes with constant covariance and the LMARCH volatility have a constant drift, as set by the CMA value for $\mu$.
Adding a DU term does not change the mean drift.

The impact of the NRC term is seen on the last 2 processes (in red and blue, with the red data below the blue ones).
Due to the short term trend following included in the NRC term, the initial NRC contribution is negative and decrease the drift as set by the CMA.
At the scale of a few years, the mean reversion dominates and increases the drift.
Then, after a few simulation years, the long term drift is set by the CMA values because the transient effects induced by the initial market conditions subside.
Adding the DU term has no effect on the mean drift, in agreement with the simple model used in Appendix~\ref{annex:toy_model_drift_uncertainty}.

\subsection{Standard deviation for the wealth}
\begin{figure}[ht]
	\centering
	\includegraphics[width=0.8\figureWidth]{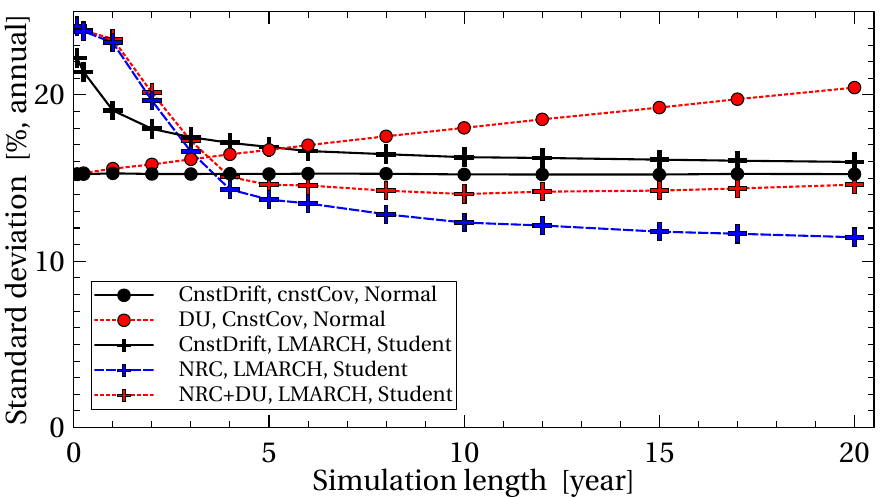}
	\caption{The standard deviation for different processes, as function of the simulation length.}
	\label{fig:wealthStdDev}
\end{figure}
The graph~\ref{fig:wealthStdDev} shows the standard deviation of the simulated wealth (vertical axis) as function of the simulation length $\DT$ (horizontal axis).
The standard deviation has been annualized, so that the leading $\sqrt{\DT}$ scaling is removed.
The base line process with constant drift, constant volatility and normal innovation  has no dependency from the past, hence the constant standard deviation, at the level sets by the CMA.
The same process but with a DU term on the drift shows clearly the linear increase of the standard deviation with the simulation length.
This larger uncertainty on the wealth is due to the uncertainty on the CMA drift, as captured by the DU term.

The LMARCH process starts with a higher value due to the crisis, and subside to the long term value as set by the CMA values.
Because the ARCH process has a long memory, the cross-over takes a few years.
Adding a NRC in the drift increase the standard deviation at short term at this starting date, due to the large value for the NRC term.
As expected, the NRC term decreases the long term standard deviation, with the difference controlled by the magnitude in the NRC term.
Intuitively, a large long term deviation from the expected drift is likely to induce a correction, hence a smaller standard deviation.
Finally, adding the NRC and DU terms increase the standard deviation linearly, as derived with a simple model in Appendix~\ref{annex:toy_model_drift_uncertainty}.
But the increase is tamed by the NRC term, which creates a pull back effect for large DU random drift.

\subsection{VaR at a 5\% level}
\begin{figure}[ht]
	\centering
	\includegraphics[width=0.8\figureWidth]{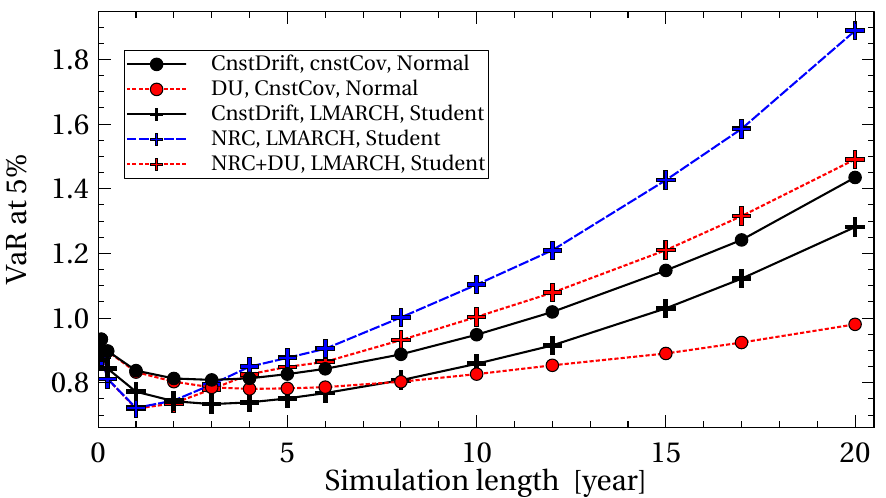}
	\caption{The 5\% relative VaR for different processes, as function of the simulation length $\DT$.
	}
	\label{fig:wealthVaR_005}
\end{figure}
The 5\% quantiles of the terminal wealth after $\DT$, equivalent to a VaR at 95\% for the losses, measure the level of wealth that is not reached in 5\% of the case.
This is a common measure of risk in portfolio management, and is reported in Fig.~\ref{fig:wealthVaR_005} as function of the simulation time $\DT$.
On this figure, a higher level means less risk, since larger values for the wealth are obtained with a 95\% probability.

Overall, the quantiles decrease during the first years, meaning more risk.
This is due to the diffusion, that dominates the drift for small $\DT$, as explained in  Sec.~\ref{sec:coreScaling}.
The details of the process matter for the 5\% VaR, and the LMARCH volatility is higher due to the Covid crisis, leading to a larger risk up to a few years.
For large $\DT$, the drift dominates, leading to the steady increases of the quantiles, therefore a diminishing risk.
In between, a cross-over takes place at the scale of a few years, in this case between 3 to 7 years.
The CMA value used for this asset are $\mu$ = 8.9\% and $\sigma$ = 16.6\%, leading to a cross-over time $\crossOverTime \simeq$ 3.5 year.
This figure gives a clear illustration of the impact of the random walk scaling ($\DT$ for the drift versus $\sqrt{\DT}$ for the volatility) and the induced cross-over.
Notice also the fairly long time needed to have the 5\% VaR at the initial value of the portfolio (depending on the process, between 8 to 20 years).

The NRC and DU terms alter the drift, hence the different long term behaviors.
As can be expected, the LMARCH volatility and DU term increase the risk, while the NRC term reduces the risk.

\subsection{Lower tail risks}
\begin{figure}[ht]
	\centering
	\includegraphics[width=0.8\figureWidth]{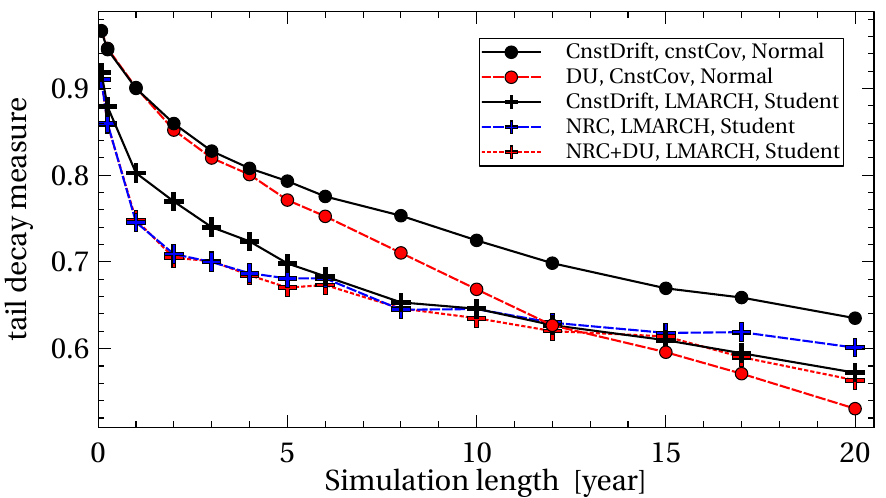}
	\caption{The VaR ratio VaR(0.01)/VaR(0.05) for different processes, as function of the simulation length $\DT$.}
	\label{fig:wealthLowerTailVaRRatio}
\end{figure}
The more extreme risks in the lower tail can be measured by VaR, say at the 1\% level.
An interesting measure is the ratio VaR(0.01)/VaR(0.05) which quantify ``how fat is the tail''.
A low value means that VaR(0.01) is low compared to VaR(0.05), namely larger extreme risks.
The figure \ref{fig:wealthLowerTailVaRRatio} shows this VaR ratio, low values mean larger tails, or more extreme risks.
The overall decreasing values with $\DT$ mean larger tails for increasing investment horizons.
Then, a LMARCH volatility and a DU term both increase the extreme risks, due to the increased probability of lower wealth.
In the other direction, the usual random walk process has the smallest tail.
This is likely underestimating the extreme risk, since all modifications of this process produce heavier tails.

\FloatBarrier
\section{Conclusions}
Multivariate long term models for the market evolution are required for a variety of long term financial plannings, like for pension funds, social insurances, or in wealth management.
The simplest model is a multivariate normal process, but the stylized facts accumulated over the last 40 years point toward more complex models.
A critical issue is the validation of such models, since the long time horizons preclude using a large enough effective sample in order to evaluate significant statistics.
The validation step is why the construction of models beyond the usual normal process has been largely ignored in the academic literature.
Yet, such models are used anyway in the industry, because plannings and validations need to be done in order to check the adequacy of an investment strategy toward the desired long term goals.
Hence, it is better to construct a model incorporating today's knowledge, even though a validation cannot be done with all the desired rigor.
The present contribution is a step in this direction, trying at best to extract long term empirical information from market data in order to validate a process.
Clearly, the exercise is fundamentally limited, yet the present contribution is an honest attempt to improve over a simple normal random walk commonly used today.

Backed by several empirical statistics aimed to extract at best statistical information at long time scales from the financial markets, several additions are made to the core normal random walk with respect to the drifts, covariance matrix and the distribution for the innovations.
Compared to the basic normal random walk, the salient outcome of using the proposed process is the increased risk.
Most components push in this direction: the drift uncertainty, the dynamic volatility, and the skewed and fat-tailed innovations.
Only the NRC term provides for a stabilization.
Another difference is the dependencies on the recent market states introduced by the NRC and the dynamic covariance, making for a different mean and risk over a few years.
Given this picture, it is possible that many long term investment plans, say pension funds, underestimate their tail risks.

\newpage
\section*{Acknowledgments}
The author is indebted to Alexandra Tchalakian for many discussions and for important contributions in the implementation of the present model and statistics.

This work was part of a larger project made with the bank J.P.Morgan, and this paper benefited from the many comments made by the project participants.

\newpage
\appendix
\renewcommand{\thesection}{\Alph{section}}
\renewcommand{\theequation}{\thesection.\arabic{equation}}

\section{A simple model with drift uncertainty}
\label{sec:drift_uncertainty}
\setcounter{equation}{0}
\label{annex:toy_model_drift_uncertainty}
Let us assume that the forecast for the drift $\mu$ is a random variable with distribution
\begin{equation}
	p(\mu) = N(\overline{\mu}, \sigma_\mu^2)
\end{equation}
where $\overline{\mu}$ is the value provided by the CMA.
The uncertainty over this value is controlled by $\sigma_\mu$.

The simplest model for the logarithmic return $r$ is a normal distribution
\begin{equation}
	p(r|\mu) = N(\DT\,\mu, \DT\,\sigma_r^2)
\end{equation}
where $\DT$ is the desired time interval up to the terminal value.
We want to evaluate the terminal return distribution, taking into account the uncertainty on the drift
\begin{equation}
	\label{eq:terminalDistribution}
	p(r) = \int d\mu \,p(r|\mu) \,p(\mu).
\end{equation}
Because both distributions are normal, the computation can be done analytically, and $p(r)$ is a normal distribution with annualized variance $\DT\,\sigma_{\text{eff}, r}^2$.
The result at leading order for the annualized standard deviation is
\begin{equation}
	\label{eq:effectiveVol}
	\sigma_{\text{eff}, r} \simeq \sigma_r \left(1 + \frac{1}{2}\,\DT\,\left(\frac{\sigma_\mu}{\sigma_r}\right)^2\right).
\end{equation}
Due to the linear scaling of the drift, the correction term for the standard deviation is also linear in $\DT$.
Let us emphasize that the volatility at time horizon $\DT$ is $\sqrt{\DT}\sigma_{\text{eff}, r}$ hence the short term leading scaling is in $\DT^{1/2}$ but the long term correction grows as $\DT^{3/2}$.

In order to see if the correction is quantitatively important, a model for the uncertainty on $\mu$ should be established, essentially relating $\sigma_\mu$ to the other quantities.
Let us assume that the model is calibrated using historical data available over a time span $\DT_\text{cal}$ (and assuming that distributions are stationary).
The realized return is one draw from the distribution $N(\DT_\text{cal} \overline{\mu}, \DT_\text{cal}\sigma_r^2)$, i.e.
\begin{equation*}
	r = \DT_\text{cal} \,\overline{\mu} + \sqrt{\DT_\text{cal}}\,\sigma_r \,\epsilon_i,
\end{equation*}
with $\epsilon_i \sim N(0, 1)$.
The historical estimation for the annualized drift is $\hat{\mu} = r/\DT_\text{cal}$, or
\begin{equation}
	\label{eq:drift_uncertainty}
	\hat{\mu} = \overline{\mu} + \frac{\sigma_r}{\sqrt{\DT_\text{cal}}} \,\epsilon_i.
\end{equation}
Therefore, the uncertainty on the estimated value is
\begin{equation}
	\sigma_\mu = \frac{\sigma_r}{\sqrt{\DT_\text{cal}}}.
\end{equation}
Inserting this relation in \eqref{eq:effectiveVol} gives
\begin{equation}
	\sigma_{\text{eff}, r} \simeq \sigma_r \left(1 + \frac{1}{2}\,\frac{\DT}{\DT_\text{cal}}\right).
\end{equation}
Typically, a few decades of historical data are available, say 20 to 40 years.
Taking say 25 years of historical data for the calibration leads to
\begin{equation}
	\sigma_\mu \approx \frac{1}{5} \,\sigma_r,
\end{equation}
which is a pretty small uncertainty.
With a target simulation time of a few decades, say 20 to 30 years, the correction term is not negligible but also not dominant.
In the range of $\DT$ amenable for statistics, essentially up to 1 year, the correction is small.

\section{Non-Central Student-t: random draws and evaluation of $\theta(\nu)$}
\label{appendix:non-central-student}
\setcounter{equation}{0}

The algorithm presented in \citep{Embrechts.2015} is modified as follow.
A random variable $w$ is introduced, with $\nu/w \sim \chi^2(\nu)$ and $\nu$ the degrees of freedom.
A convenient function is
\begin{equation}
	\label{def:theta}
	\theta(\nu) = \frac{\var{\sqrt{w}}}{\E{w}} = 1 - \frac{ (\E{\sqrt{w}})^2 }{\E{w}}
\end{equation}
which measure in relative term the variance of the mixing variable $\sqrt{w}$.
The non-centrality parameter is $\bm{\gamma}$, and a matrix $\bm{\chi}$ is defined as
\begin{align}
	\bm{\chi} = \id + \theta \,\bm{\gamma}\cdot\bm{\gamma}'
\end{align}
The matrix $\bm{\chi}$ is symmetric and positive definite, hence all its eigenvalues are real and positive.
Consequently, the inverse square root of the matrix $\bm{\chi}$ is well defined, and can be evaluated from the eigenvalues and eigenvectors.
Finally, because the inverse and the square root are well defined, the relation
\begin{align}
	\label{eq:chi_product}
	\bm{\chi}^{-1/2} \cdot \bm{\chi} \cdot \bm{\chi}^{-1/2} = \id
\end{align}
holds.

\begin{figure}
	\centering
	\includegraphics[width=0.6\linewidth]{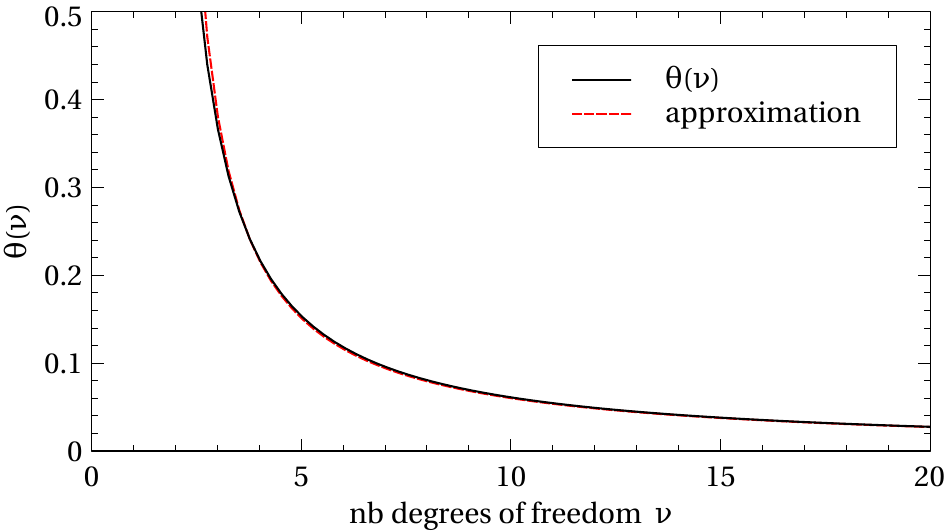}
	\caption{The function $\theta$ and its approximation versus $\nu$.}
	\label{fig:theta}
\end{figure}

The multivariate non-central Student random generator for the variable $\bm{\epsilon}$ is defined by
\begin{align}
	\label{eq:random_non_central_student_innovation}
	\bm{\epsilon} & = \bm{\chi}^{-1/2} \left\{\frac{\sqrt{w} - \E{\sqrt{w}}}{\sqrt{\E{w}}} \cdot\bm{\gamma} + \frac{\sqrt{w}}{\sqrt{\E{w}}} \bm{Z} \right\}
\end{align}
where $\bm{Z}$ is a vector of independent normal variable with $z_\alpha \sim N(0,1)$, and $\bm{\epsilon}$ is a random vector of innovations.
The explicit expressions for $\E{\sqrt{w}}$ and $\E{w}$ are given below in Eq.~\ref{eq:momentsForW}.
Notice that with the definitions~\ref{eq:random_non_central_student_innovation}, both fractions depending on $w$ are independent of the scale selected for $w$, namely the same random variables $\bm{\epsilon}$ are obtained when using  $\alpha/w \sim \chi^2(\nu)$ regardless of $0 < \alpha < \infty$.
The usual convention is to take $\alpha = \nu$.
With this specifications, the vector of random innovations $\bm{\epsilon}$ has a non-central Student distribution with $\E{\bm{\epsilon}} = 0$ and $\var{\bm{\epsilon}} = \id$ (where the relation \ref{eq:chi_product} is used in the derivation).
The non-central Student random number generation involves the expectations for $\sqrt{w}$ and $w$, where $\nu/w$ is distributed according to a chi-square distribution.
In the scalar case, and for a slightly different specification, these expectations are available in wikipedia \citep{wikipedia_nonCentralT}.
By identification, the desired expectations are obtained.
The wikipedia specification for the scalar random variate $x$ is
\begin{equation}
	x = \mu + \sqrt{w}\,\gamma + \sqrt{w} \,z.
\end{equation}
For these specifications, the following expression are obtained for the mean and variance
\begin{subequations}
	\begin{align}
		\E{x} & = \mu + \E{\sqrt{w}}\,\gamma, \\
		\var{x} & = \E{w} \cdot 1 + \var{\sqrt{w}} \gamma^2.
	\end{align}
\end{subequations}
Comparing with the explicit expressions given in wikipedia, the desired expectations are obtained
\begin{subequations}
	\label{eq:momentsForW}
	\begin{align}
		\E{\sqrt{w}} & \cong \left(1 - \frac{3}{4\nu - 1} \right)^{-1}, \\
		\E{w} & = \frac{\nu}{\nu - 2}.
	\end{align}
\end{subequations}
From both expressions, the value of $\theta = \theta(\nu)$ is obtained.
An asymptotic expansion in $1/\nu$ leads to $\theta \simeq 1/(2\nu)$, while a very good approximation in the domain of interest is $\theta \simeq 1/(2\,(\nu - 1.7))$.
The value of $\theta$ and the approximation are shown in figure \ref{fig:theta}.
Essentially, in the region of interest with $\nu \simeq 6$ to 10, $\theta(\nu)$ is small with $\theta \simeq$ 0.12 to 0.05.

\section{Long term expectations}
\label{appendix:longTermExpectations}
This section focuses on the long term expectations of the process, in particular for the drift and covariance matrix.
For this section, the multivariate process includes a constant drift (i.e. no NRC and DU term), a LMARCH volatility, and non-central Student innovations.
The stationary values are assumed to exist, namely for a sufficiently large time $t$, the values $\E{\bm{r}(t)}$ and $\E{\bm{r}(t) \cdot \bm{r}(t)'}$ converge to a given vector and matrix, respectively, that are independent of the time $t$ and of the initial conditions.
The expectations $\E{\cdot}$ are taken with respect to the random realization of the innovations distributed according to a non-central student distribution.

For the mean, equation \ref{eq:random_non_central_student_return} together with $\E{\bm{\epsilon}} = 0$ gives the desired result, namely
\begin{equation}
	\E{\bm{r(t)}} = \bm{\mu}  \hspace{2em}\forall t.
\end{equation}
This result is stronger than needed since valid for all times.

The covariance $\bm{\Sigma}(t)$ is defined in Eq.~\ref{eq:lmarch_linear} and \ref{eq:lmarch_affine}, repeated here
\begin{subequations}
	\begin{align}
		&	\bm{\Sigma}_\text{LM-ARCH, linear}(t)
		= \sum_{0 \leq k < k_\text{max}}  w(k,  \dt) ~\left(\bm{r}(t - k\,\dt) - \bm{\mu}\right) \cdot \left(\bm{r}(t - k\,\dt)  - \bm{\mu}\right)' \label{eq:lmarch_affine_one} \\
		&	\bm{\Sigma}(t) = \wInfty \cdot \bm{\Sigma}_\CMA + (1 - \wInfty) \cdot \bm{\Sigma}_\text{LM-ARCH, linear}(t).  \label{eq:lmarch_affine_two}
	\end{align}
\end{subequations}
and where the time scale $\dt$ is implicit in $\bm{r}$ and $\bm{\Sigma}_\text{LM-ARCH, linear}$.
For times $t$ larger than the memory of the LM-ARCH process, namely $t \gg t_0 + k_\text{max} \,\dt$ where $t_0$ is the start time of the simulations, all the vectors $\bm{r}$ used to compute $\bm{\Sigma}_\text{LM-ARCH, linear}$ are random and drawn from a non-central Student distribution with covariance $\bm{\Sigma}(t)$.
Assuming a fixed value $\overline{\bm{\Sigma}}$ for the expectation of $\bm{\Sigma}(t)$ for large enough $t$, the linear covariance is
\begin{align}
	\E{\Sigma_\text{LM-ARCH, linear}(t)} & = \sum_{0 \leq k < k_\text{max}}  w(k,  \dt) \,\E{\bm{\Sigma}(t - k\,\dt)} \nonumber\\
	& = \sum_{0 \leq k < k_\text{max}}  w(k,  \dt) \,\overline{\bm{\Sigma}} \nonumber\\
	& = \overline{\bm{\Sigma}}
\end{align}
where the normalization property $\sum_k w(k,  \dt) = 1$ of the LM-ARCH weights has been used.
Taking the expectation of \ref{eq:lmarch_affine_two} leads to
\begin{equation}
	\label{eq:E_Sigma}
	\overline{\bm{\Sigma}} = \wInfty \cdot \bm{\Sigma}_\CMA + (1 - \wInfty) \cdot \overline{\bm{\Sigma}}
\end{equation}
with solution
\begin{equation}
	\overline{\bm{\Sigma}} = \bm{\Sigma}_\CMA
\end{equation}
for $\wInfty > 0$.
This shows that for a (non-central) student distribution, the long term expectation of the covariance is the CMA covariance matrix, regardless of the distribution for $\bm{\epsilon}$ as parametrized by $\nu$ and $\bm{\gamma}$.



\bibliographystyle{plainnat}
\bibliography{../../Shared/LaTeX/biblioUniverse}

\end{document}